\documentclass[
    reprint,
    superscriptaddress,
    amsmath,amssymb,
    aps,
    prx,
    longbibliography,
]{revtex4-2}
\usepackage{braket}
\usepackage{graphicx}
\usepackage{amsmath}
\usepackage{amssymb}
\usepackage{amsbsy}
\usepackage{algorithm}
\usepackage{algpseudocode}
\usepackage{bbm}
\usepackage{multirow}
\usepackage{xcolor}
\usepackage[margin=0.75in]{geometry}

\usepackage{subcaption}
\usepackage{booktabs}
\usepackage[english]{babel}
\usepackage{tikz}
\usetikzlibrary{matrix,shapes,calc}

\usepackage[colorlinks=true,urlcolor=blue,linkcolor=blue,citecolor=blue]{hyperref}

\DeclareMathOperator*{\argmin}{arg\,min}

\newcommand{\skiptext}[1]{}

\newcommand{\x}{\mathbf{x}}

\newcommand{\Eq}[1]{Eq.(\ref{#1})}
\newcommand{\Fig}[1]{Fig.\ref{#1}}
\newcommand{\Tab}[1]{Tab.\ref{#1}}

\definecolor{LBlue}{rgb}{0,0.34,0.45}

\tikzstyle{block} = [draw, fill=white, rectangle, 
    minimum height=0.8cm, minimum width=0.8cm, inner sep=.5pt]
\tikzstyle{big_tensor} = [draw, fill=white, rectangle, 
    minimum height=0.8cm, minimum width=6.8cm, inner sep=.5pt]
\tikzstyle{blob} = [draw, fill=white, circle, 
    minimum height=0.55cm, minimum width=0.55cm, inner sep=.5pt]

\begin{document}

\title{Boosting Binomial Exotic Option Pricing with Tensor Networks}

\author{Maarten van Damme}
\affiliation{SandboxAQ, Palo Alto, CA, USA}

\author{Rishi Sreedhar}
\affiliation{SandboxAQ, Palo Alto, CA, USA}

\author{Martin Ganahl}
\email{martin.ganahl@sandboxaq.com}
\affiliation{SandboxAQ, Palo Alto, CA, USA}

\date{\today}

\begin{abstract}
    Pricing of exotic financial derivatives, such as Asian and multi-asset American basket options, poses significant challenges for standard numerical methods such as binomial trees or Monte Carlo methods. While the former often scales exponentially with the parameters of interest, the latter often requires expensive simulations to obtain sufficient statistical convergence. This work combines the binomial pricing method for options with tensor network techniques, specifically Matrix Product States (MPS), to overcome these challenges. Our proposed methods scale linearly with the parameters of interest and significantly reduce the computational complexity of pricing exotics compared to conventional methods. For Asian options, we present two methods: a tensor train cross approximation-based method for pricing, and a variational pricing method using MPS, which provides a stringent lower bound on option prices. For multi-asset American basket options, we combine the decoupled trees technique with the tensor train cross approximation to efficiently handle baskets of up to $m = 8$ correlated assets. All approaches scale linearly in the number of discretization steps $N$ for Asian options, and the number of assets $m$ for multi-asset options. Our numerical experiments underscore the high potential of tensor network methods as highly efficient simulation and optimization tools for financial engineering.
\end{abstract}

\maketitle
\section{Introduction}

Options are financial derivatives that give the holder the right, but not the obligation, to buy (call option) or sell (put option) an underlying asset at a predetermined price within a specific time frame \cite{hull2021}. These versatile instruments allow investors to hedge against market volatility, speculate on price movements, and construct complex trading strategies. Accurate pricing of options is essential for fair valuation, informed decision-making, and efficient markets. Hence, option pricing is a fundamental problem in quantitative finance, crucial for risk management, investment strategies, and financial market stability.

Various techniques have been developed to price options, ranging from analytical methods like the Black-Scholes model \cite{black1973} to numerical approaches such as binomial trees \cite{cox1979, rendleman1979} and Monte Carlo simulations. While these methods have proven effective for standard options, they face significant challenges when applied to path-dependent options (e.g., Asian options) or multi-asset options (e.g., basket options), the primary obstacle being the curse of dimensionality where as the number of time steps or underlying assets increases, the computational complexity grows exponentially [see below]. This limitation has spurred ongoing research into more efficient pricing methods for complex options, particularly those involving multiple assets or path-dependent payoffs \cite{Buchen2012}, including machine learning techniques \cite{Ruf2019}.

Tensor network methods are a class of novel algorithmic approaches that are receiving an increasing amount of attention for applications in finance \cite{Glau2020, SAKURAI2024, kobayashi2024, Patel2022, Kastoryano2022, Cassel2022, antonov_alternatives_2021, xu_tensor-train_2021, mugel_use_2020, mugel_dynamic_2022, patel_quantum-inspired_2022, patel_application_2024}. Tensor networks \cite{perez-garcia_matrix_2006, verstraete_matrix_2008, schollwoeck_density-matrix_2011, evenbly_algorithms_2009, chan_matrix_2016,  evenbly_practical_2022,orus_practical_2013, cirac_matrix_2021} are a class of highly efficient data structures to store and manipulate high-dimensional arrays. Originally explored as computational tools for solving the quantum many-body problem \cite{wilson_renormalization_1975, affleck_rigorous_1987, rommer_class_1997}, they are today considered the gold standard methods for approximating ground-states of one-dimensional \cite{white_density_1992, white_density-matrix_1993}, two-dimensional \cite{verstraete_renormalization_2004} and quantum-critical \cite{vidal_entanglement_2007, vidal_class_2008, evenbly_algorithms_2009} systems, for computing quantum-chemical properties of molecules \cite{legeza_qc-dmrg_2003, white_ab_1999, chan_density_2011} and materials \cite{garcia_dynamical_2004, ganahl_chebyshev_2014, ganahl_efficient_2015, bauernfeind_fork_2017, menczer_two-dimensional_2024, wolf_chebyshev_2014, wolf_solving_2014}, and are rapidly expanding into the machine learning and computer science domain \cite{dolgov_approximation_2019, DOLGOV2020, OSELEDETS2010, glasser_expressive_2019, glasser_probabilistic_2020, Cichocki2016p1, Cichocki2017p2, goesmann_tensor_2020, ali_anomaly_2024, wang_anomaly_2020, stoudenmire_learning_2018, wang_principal_2018, lu_tensor_2021, liu_tensor_2023, konstantinidis_bayesian_nodate, kirstein_tensor-train_2022, izmailov_scalable_2018, han_unsupervised_2018, peng_generative_2023, strashko_generalization_2022}.
%
%These techniques have been instrumental in simulating quantum many-body systems \cite{Ors2019}, solving complex optimization problems \cite{Cichocki2016p1, Cichocki2017p2}, and simplifying machine learning models \cite{Roberts2019, Efthymiou2019, Sengupta2022}. 
The ability of tensor networks to efficiently represent and manipulate high-dimensional data makes them a promising candidate for addressing the challenges in financial engineering in general and option pricing in particular \cite{Glau2020, SAKURAI2024, kobayashi2024, Patel2022, Kastoryano2022, Cassel2022, antonov_alternatives_2021, xu_tensor-train_2021, mugel_use_2020, mugel_dynamic_2022}. 
%This paper explores a novel approach to pricing path-dependent and multi-asset options by adapting tensor network techniques, specifically the Tensor Train Cross (TTCross) algorithm \cite{OSELEDETS2010}, to financial modeling. By leveraging these quantum-inspired methods, we aim to provide a simple yet powerful framework for managing the curse of dimensionality in option pricing, potentially opening new avenues for interdisciplinary research between physics and finance.
This work introduces novel tensor network approaches for pricing exotic options. The algorithms we present are based on the binomial pricing method, a simple, robust, and widely used method for pricing options \cite{cox1979, rendleman1979}. While being mainly used in practice to price vanilla derivatives, binomial pricing suffers from an exponential scaling for many exotic options, rendering its application for such options often impractical or even impossible. The novelty of our proposed methods lies in overcoming this exponential scaling and making binomial pricing applicable to these problems.

\section{Options}
\label{sec:options_in_finance}

The two main classes of options are Call and Put options, which give the holder the right to buy and sell the underlying asset at the specified strike price $K$, respectively. Options offer investors and traders a versatile tool for managing risk, generating income, speculating, and diversifying their portfolios, making them an important component of modern financial markets. Simple options are typically defined by an underlying asset with asset price $S$, a strike price $K$, and a payoff function $v(S)$. The payoff function defines the value of the option at exercise, with a typical payoff function for a call option being $v(S) = \max (S-K, 0)$ (but more complicated functions are used in practice as well, see below). 

Options are usually classified into {\it vanilla} and {\it exotic} options, depending on how widely traded they are. Vanilla options have relatively simple payoff structures and are based on standard terms and conditions. Exotic options, on the other hand, are more complex financial instruments with non-standard payoff structures or customized features that differ from plain vanilla options. They are designed to meet specific risk-return profiles or to capitalize on particular market conditions. Options are further categorized into American or European style options, depending on whether they are exercisable at any time before or only at the expiration date, respectively.

A key challenge when dealing with options is obtaining accurate estimates of their value. Given a strike price $K$, an initial asset value $S_0$ at time $t = 0$, a terminal price $S_T$ at exercise time $t = T$, and the option payoff function $v(S_T,K)$, the theoretical value $V(t, K|S_0)$ of the option is given by the Feynman-Kac formula 
\begin{equation}
\label{eq:feynman_kac}
    V(t,K|S_0) = e^{-r(T-t)}\mathbb{E}[v(S_T,K)|S_0].
\end{equation}
$\mathbb{E}[\cdot]$ denotes the expectation value over the possible paths of the asset price $S_t$ between $0<t<T$, $r$ is the continuously compounded risk neutral interest rate, and $e^{-r(T-t)}$ accounts for the present value from future cashflow. Following standard practice, we assume that assets follow standard geometric Brownian motion, i.e.
given $m$ asset values $S^i_t, i = 1,\dots,m$ at time $t$,
\begin{equation}
\label{eq:gbm_multi-asset}
    dS^i_t = S^i_t r dt + \sigma_iS_t^idW^i_t
\end{equation}
with volatilities $\sigma_i$, and $dW^i_t$ multivariate normally distributed with mean $\mathbb{E}[dW^i_t]=0\;\forall i$ and correlation matrix $\rho_{ij}(t)$ and covariance matrix $\Sigma$ with 
\begin{align}
    \rho_{ij} dt &\equiv \textrm{corr}[\frac{dS^i_t}{S^i_t},\frac{dS^j_t}{S^j_t}]=\mathbb{E}[dW^i_tdW^j_t]\label{eq:corr}\\
    \Sigma_{ij} &\equiv \sigma_i\sigma_j\rho_{ij}.\label{eq:covar}
\end{align}
In this work, we will primarily focus on the pricing of exotic options (which is usually computationally much harder than vanilla options) \cite{Buchen2012}. In particular, we focus on standard Asian and American basket options, which we briefly introduce in the following. 
%In the following, we will mostly focus on call options, but all results of this manuscript apply equally well to put options. 

\textbf{Asian options} are characterized by a payoff function $v_{\text{A}}$ which depends on the average price of the underlying asset over a predefined averaging period and sampling frequency, i.e. 
\begin{equation}
\label{eq:asian_option_payoff}
    \begin{split}
        v_{\text{A}} &= \text{max}(\langle S_{T} \rangle - K,0), \\
    \end{split}
\end{equation}
with $\langle S_{T} \rangle$ denoting the time-averaged asset value along a given asset trajectory $S_t,\;  0\leq t \leq T$. While we focus on Asian Call options in the following, all algorithms presented here and below are straightforwardly extended to Put options.
The path dependence of the payoff function makes Asian options, in general, more complicated to price. For example, standard binomial pricing methods (see below), used widely to price vanilla options, in general become exponentially costly, and are not readily applicable to this case. For this work, we will focus on the case of arithmetic averaging $\langle S\rangle = \int_{t=0}^{T} S_t dt$ (for geometric averaging under standard Brownian motion of $S_t$, certain Asian options can be priced analytically). Conventional methods used in this case are Monte Carlo sampling or finite difference methods for integrating the associated equations of motion of the expected value $V(t, K|S_0)$ \cite{Jasra2011, Duffy2006}, which can become computationally expensive. 

\textbf{European and American basket options} are extensions of standard options from a single underlying asset $S$ to $m>1$ underlying assets $S^1, S^2, \dots S^m$ (a "basket" ). Typical payoff functions of put options are functions of min, max, or mean basket value, i.e. 
%function of the value of all the assets in the basket, making them harder to price efficiently. Examples include min, max, or weighted average baskets, where the payoff is the least valued asset price, the most valued asset price, or a weighted average of all asset prices in the basket relative to the strike price, respectively. That is,
\begin{equation}
\label{eq:amr_basket_option_payoff}
    \begin{split}
        v_{\text{max}} &= \text{max}(K - \text{max}(S^1, \dots,S^m),0) \\
        v_{\text{min}} &= \text{max}(K - \text{min}(S^1, \dots,S^m),0) \\
        v_{\text{avg}} &= \text{max}(K - \text{avg}(S^1, \dots,S^m),0).
    \end{split}
\end{equation}
Standard pricing methods like binomial pricing scale exponentially in the number of assets $m$, limiting their applicability to small baskets. Again, Monte Carlo techniques 
%and their variants, like least square Monte Carlo, 
are typically used to price these derivatives. Similar to the Asian case, getting accurate results can often require expensive, time-consuming computer simulations \cite{Bouchard2012}.

\section{Tensor networks and matrix product states}
\label{sec:ttx}
Tensor networks are a class of highly efficient data structures to store and manipulate certain classes of high-dimensional arrays. In this work, we focus on the so-called matrix product state (MPS) ansatz (more recently also known as tensor train \cite{OSELEDETS2010}), the most prominent and successful example of a tensor network.

Consider an $N$ dimensional array, or tensor, $M_{x_1 x_2\dots x_N}$, with indices $x_i = \{0,1,\dots,d_i\}$ labeling the tensor elements. The tensor $M_{x_1 x_2\dots x_N}$ is said to be in MPS format if it has the form
\begin{align}
    M_{x_1 x_2\dots x_N}&= \sum_{\{\alpha_1,\dots,\alpha_{N-1}\}} A_{\alpha_1}^{x_1} A_{\alpha_1\alpha_2}^{x_2} \dots A_{\alpha_{N-1}}^{x_N}\label{eq:mps}\\
    &= A_1^{x_1}A_2^{x_2} \dots A_N^{x_N}.\nonumber
\end{align}
where we use the second line as a shorthand for \Eq{eq:mps}.
Here, $\{A^{x_k}_{\alpha_{k-1}\alpha_k}\}$ are order-3 tensors with indices $(x_k, \alpha_{k-1},\alpha_k)$ of dimensions $(d_k, D_{k-1}, D_k)$, respectively, and $A^{x_1}_{\alpha_1}$ and $A^{x_N}_{\alpha_{N-1}}$ are order-2 tensors. The maximum value $D\equiv \max[D_1, \dots, D_{N-1}]$ is known as the {\it bond dimension} of the MPS. 

A key feature of \Eq{eq:mps} is the {\it linearly} scaling memory requirement with increasing number $N$ of axes of the MPS (at a fixed bond dimension). This is in contrast to the exponential memory scaling for generic tensors, often referred to as the {\it curse of dimensionality}. The bond dimension $D$ is a hyperparameter of the ansatz, with larger values of $D$ typically giving a more expressive ansatz. These features, in combination with efficient algorithms to manipulate them, make MPS ideally suited for tackling high-dimensional optimization problems, bypassing the usual curse of dimensionality. 
Note that in the limit of exponentially growing $D \sim \exp(N)$, the MPS ansatz recovers the full space of order-$N$ tensors. 

In the following, we use MPS to approximate and optimize multivariate functions $f(x_1, \dots, x_N)$ with $N$ discrete variables.
Without loss of generality we can assume $x_i$ to take integer values, such that $f(x_1, \dots x_N)$ can naturally be represented as a tensor, i.e. 
\begin{align}
    f_{x_1x_2\dots x_N} &\equiv f(x_1,x_2,\dots,x_N) \nonumber\\
    x_i \in &\{0,1,\dots,d_i\}\\
    \x \equiv& (x_1, \dots, x_N).
\end{align}
From an MPS representation of $f(\x)$, complex quantities like high-dimensional sums, or partition functions and high-order moments of probability distributions can be computed in a time linear in $N$, using well-known standard tensor network contraction methods \cite{evenbly_practical_2022, bridgeman_hand-waving_2017}.
We will use these methods to our advantage when pricing complex financial derivatives, specifically path-dependent Call or Put options, and high-dimensional European and American basket options, for which only either exponentially scaling methods or Monte Carlo approaches can be used.

A key step in the approach is obtaining an accurate MPS approximation of the discrete function $f(\x)$ 
\cite{dolgov_approximation_2019,peng_generative_2023, OSELEDETS2010, steinlechner_riemannian_2016}. 
In this work, we employ a standard vectorized, parallel tensor-train cross (TTCross) implementation \cite{DOLGOV2020} 
build on top of the TensorNetwork package \cite{Roberts2019}.

\section{Binomial pricing of exotic options with tensor networks}
\label{sec:binomial_model}

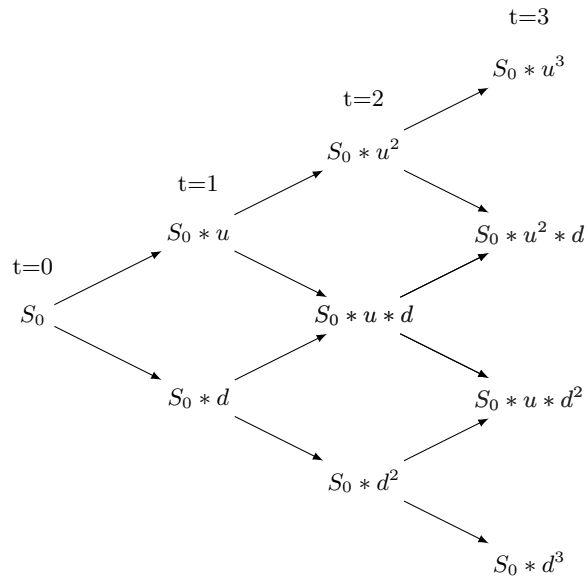
\begin{figure}
    \centering
    \begin{tikzpicture}[grow=right, edge from parent/.style={draw,-latex}, label distance = 0.2cm,
      level 1/.style = {level distance=22mm, sibling distance=22mm},
      level 2/.style = {level distance=22mm, sibling distance=22mm},
      level 3/.style = {level distance=22mm, sibling distance=22mm},
      ]
      \node[label=90:{t=0}] {$S_0$}
      child {node {$S_0*d$}
        child {node {$S_0*d^2$}
          child {node {$S_0*d^3$}}
          child {node[opacity=.5] {$S_0*u*d^2$}}
        }
        child {node[opacity=.75] {$S_0*u*d$}
          child {node[opacity=.5] {$S_0*u*d^2$}}
          child {node[opacity=.5] {$S_0*u^2*d$}}
        }
      }
      child {node[label=90:{t=1}] {$S_0*u$}
        child {node[opacity=.75] {$S_0*u*d$}
          child {node[opacity=.5] {$S_0*u*d^2$}}
          child {node[opacity=.5] {$S_0*u^2*d$}}
        }
        child {node[label=90:{t=2}] {$S_0*u^2$}
          child {node[opacity=.5] {$S_0*u^2*d$}}
          child {node[label=90:{t=3}] {$S_0*u^3$}}
        }
      };
    \end{tikzpicture}
    \caption{A Three-period binomial model describing the asset price dynamics as a recombinant tree.}
    \label{fig:3pBinomialTree}
\end{figure}

The binomial pricing model is widely used to price real-world options for financial applications. In the binomial pricing method, the lifetime $T$ of a derivative is discretized into $N+1$ discrete, equally spaced time points with separation $\Delta t=\frac{T}{N}$. The continuous evolution of the price $S_t$ of an underlying asset is then approximated by a discrete random walk $S_{t_k}$, $k = 0\dots N$. During each  discrete step, the asset price $S_{t_k}$ can move up or down with a probability $p_u$ and $p_d = 1-p_u$, respectively, i.e. 
\begin{align}
S_{t_{k+1}} = 
    \begin{cases}
        u S_{t_k} \textrm{with probability} \;p_u\\
        d S_{t_k} \textrm{with probability} \;p_d = 1-p_u,
    \end{cases}
\end{align}
as shown in \Fig{fig:3pBinomialTree}.
The parameters $u, d$ and $p_u, p_d$ are usually obtained assuming geometric Brownian motion of the asset price $S_t$ and requiring convergence of the discrete evolution of the derivative value to the Black-Scholes-Merton results in the continuous limit $\Delta t \rightarrow 0$. Two widely used schemes, which we also use in this work, are the Cox, Ross, and Rubinstein \cite{cox1979} (CRR) and the Rendleman-Bartter \cite{rendleman1979} (RB) schemes with values shown in \Tab{tab:CRR-RB}.
\begin{center}
    \begin{table}
    \caption{Binomial pricing parameters for Cox, Ross, and Rubinstein and Rendelman-Bartter schemes. }
    \begin{tabular}{ |c||c|c|c| } 
     \hline
     &$p_u$& $u$ & $d$ \\ 
     \hline\hline
     CRR &$\frac{e^{r\Delta t} - d}{u - d}$ &$e^{\sigma\sqrt{\Delta t}}$&$\frac{1}{u}$\\ 
     \hline
     RB & $\frac{1}{2}$ & $e^{ \left( r - \frac{1}{2} \sigma^2\right)\Delta t +\sigma\sqrt{\Delta t}}$ & $e^{ \left( r - \frac{1}{2} \sigma^2 \right)\Delta t -\sigma\sqrt{\Delta t}}$\\ 
     \hline
    \end{tabular}
    \label{tab:CRR-RB}
    \end{table}
\end{center}
%\begin{align}
%\label{eq:crr_parameters}
%        &u_{\text{CRR}} = e^{\sigma\sqrt{\Delta t}}, &u_{\text{RB}} = e^{ \left( r - \frac{1}{2} \sigma^2\right)\Delta t +\sigma\sqrt{\Delta t}} \\
%        &d_{\text{CRR}} = \frac{1}{u}, &d_{\text{RB}}=e^{ \left( r - \frac{1}{2} \sigma^2 \right)\Delta t -\sigma\sqrt{\Delta t}}\\
 %       &p_{\text{CRR}}= \frac{e^{r\Delta t} - d}{u - d}, &p_{\text{RB}} = \frac{1}{2}.
%\end{align}
%with $r$ the risk-free interest rate, $\sigma$ the asset volatility, and $\Delta t = \frac{T}{N}$ the discretization time step for a contract expiring at time $t = T$. 

\subsection{Binomial Asian option pricing with matrix product states}
The binomial model is used widely in practice to price vanilla-type European and American options.
For Asian type options, however, it faces an exponential computational scaling with the number $N$ of time steps. 
For a discrete random walk $S_{t_k}$ of the asset, the Feynman-Kac formula for an Asian option with arithmetic averaging assumes the form 
\begin{equation}
\label{eq:asian_call_value_disc}
    V(t,K|S_0) = e^{-r(T-t)}\mathbb{E}[\text{max}(\frac{1}{N} \sum_{k = 1}^N S_{t_k} - K,0)]
\end{equation}
where the expectation value $\mathbb{E}[\cdot]$ is taken over all possible paths of the asset price starting at $S_0$. The path dependence of the payoff function in general leads to exponential scaling with the number of time steps $N$ in the binomial pricing method. In the following, we present two novel approaches, scaling {\it linearly} in $N$ for a fixed bond dimension $D$, to obtain accurate estimates for Asian option prices. The performance of both methods was validated numerically and compared with standard Monte Carlo calculations over different settings. For all calculations, unless otherwise specified, we used an initial asset value $S_0 = 100$, strike price $K = 100$, time period $T = 1$, risk-free interest rate $r = 0.1$, and an asset volatility $\sigma = 0.5$.

\subsubsection{Asian option pricing with TTcross}
\label{sssec:asn_ttcross_method}
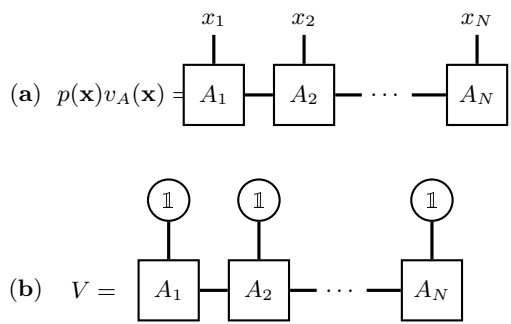
\begin{figure}
    \centering
    \begin{tikzpicture}

        % first row
        \node[] (m1) {$\mathbf{(a)}$};
        \node[right of =m1, xshift=.15cm] (0) {$ \ \ \ p(\mathbf{x})v_A(\mathbf{x}) = $};
    
        \node[block, thick, right of=0, xshift=.35cm] (a) {$A_{1}$};
            \node[above of=a] (a1) {$x_1$};
        
        \node[block, thick, right of=a, xshift=.20cm] (b) {$A_{2}$};
            \node[above of=b] (b1) {$x_2$};
        %\node[right of=b, xshift=.10cm] (d) {$\dots$};
        %\node[block, thick, right of=d, xshift=.20cm] (e) {$A_{k}$};
        %    \node[above of=e] (e1) {$x_k$};
        
        \node[right of=b, xshift=.10cm] (f) {$\dots$};
        
        \node[block, thick, right of=f, xshift=.20cm] (h) {$A_{N}$};
            \node[above of=h] (h1) {$x_N$};
        
        %\draw[black,very thick] (a) -- (b) -- (d) -- (e) -- (f) -- (h);
        \draw[black,very thick] (a) -- (b) -- (f) -- (h);
        \draw[black,very thick] (a) -- (a1);
        \draw[black,very thick] (b) -- (b1);
        %\draw[black,very thick] (e) -- (e1);
        \draw[black,very thick] (h) -- (h1);
        
        % second row
        \node[below of=m1, yshift=-0.3cm] (m2) {};
        \node[below of=m2, yshift=-0.3cm] (m3) {$\mathbf{(b)}$};
        \node[right of =m3, xshift=-.10cm] (02) {$V = $};
        
        \node[block, thick, right of=02] (a2) {$A_{1}$};
            \node[blob, thick, above of=a2, yshift=.2cm] (a12) {$\mathbbm{1}$};
        
        \node[block, thick, right of=a2, xshift=.20cm] (b2) {$A_{2}$};
            \node[blob, thick, above of=b2, yshift=.2cm] (b12) {$\mathbbm{1}$};

        \node[right of=b2, xshift=.10cm] (d2) {$\dots$};

        %\node[block, thick, right of=d2, xshift=.20cm] (e2) {$A_{k}$};
        %    \node[blob, thick, above of=e2, yshift=.2cm] (e12) {$\mathbbm{1}$};
        
        %\node[right of=e2, xshift=.10cm] (f2) {$\dots$};
        
        \node[block, thick, right of=d2, xshift=.20cm] (h2) {$A_{N}$};
            \node[blob, thick, above of=h2, yshift=.2cm] (h12) {$\mathbbm{1}$};
        
        \draw[black,very thick] (a2) -- (b2) -- (d2) -- (h2);
        \draw[black,very thick] (a2) -- (a12);
        \draw[black,very thick] (b2) -- (b12);
        \draw[black,very thick] (h2) -- (h12);

    \end{tikzpicture}       
    
    \caption{($\mathbf{a}$): Tensor network notation of a matrix product state (MPS) encoding the function $p(\mathbf{x})v_A(\mathbf{x})$. ($\mathbf{b}$): The tensor network contraction that results in a scalar that is equal to $\sum_{\mathbf{x}} p(\mathbf{x})v_A(\mathbf{x})$. This value is equal to the price of the Asian option. The contraction is between the MPS represented by blocks labelled by tensors $A^i, i \in \{ 1, \dots, N \}$, and vectors of ones $\mathbbm{1} = [1,1]^T$.}
    \label{fig:asian_mps_contraction}
\end{figure}

For an $N$-period binomial model, the asset trajectories $S_{t_k}$ can be represented as an $N$-bit binary string $ \mathbf{x} = x_1x_2\dots x_N$ with $x_k \in \{0, 1\}$, i.e.
\begin{equation}
    S_{T}(\mathbf{x}) = S_0 \Pi_{k = 1}^{N} d^{1 - x_k}u^{x_k},
\end{equation}
The probability of a path $\mathbf{x}$, the average asset price $\langle S_{T} \rangle$ and payoff $v(\mathbf{x})$ are given by
\begin{align}
    &p(\mathbf{x}) = \prod_{k = 1}^{N} p_u^{x_k}(1-p_u)^{1 - x_k},\nonumber\\
    &\langle S_{T}(\mathbf{x}) \rangle = \frac{S_0}{N} \sum_{i = 1}^N \left( \prod_{k = 1}^{i} d^{1 - x_k}u^{x_k} \right)\label{eq:asn_asset_avg}\\
    &v_A(\mathbf{x}) = \text{max} \left( \left( \left( \frac{S_0}{N} \sum_{i = 1}^N \left( \prod_{k = 1}^{i} d^{1 - x_k}u^{x_k} \right) \right) - K \right), 0 \right). \label{eq:asn_option_price_bm_traj}
\end{align}
Here, $p_u$ is the probability of an up-move of the asset price. Using the Feynman-Kac expression \Eq{eq:asian_call_value_disc}, the option price is given by
\begin{equation}
    V(t=0, K|S_0) = e^{-rT}\sum_{\mathbf{x}} p(\mathbf{x}) v_A(\mathbf{x}).\label{eq:asian-final}
\end{equation}
While $p(\x)$ has a simple product structure, the path-dependence of the payoff function $v_A(\x)$ 
renders the sum in \Eq{eq:asian-final} intractable. To deal with this sum 
we apply the TTcross method to approximate the product $p(\mathbf{x})v_A(\mathbf{x})$ 
as an MPS with tensors $\{ A^i \}$, such that
\begin{equation}
    p(\mathbf{x})v_A(\mathbf{x}) \approx A_1^{x_1} A_2^{x_2} \dots A_N^{x_N},
\end{equation}
and then use standard tensor network contraction methods for approximating the sum in \Eq{eq:asian-final}, see \Fig{fig:asian_mps_contraction} (a) and (b).

\begin{figure}
    \centering
    \includegraphics[width=0.95\linewidth]{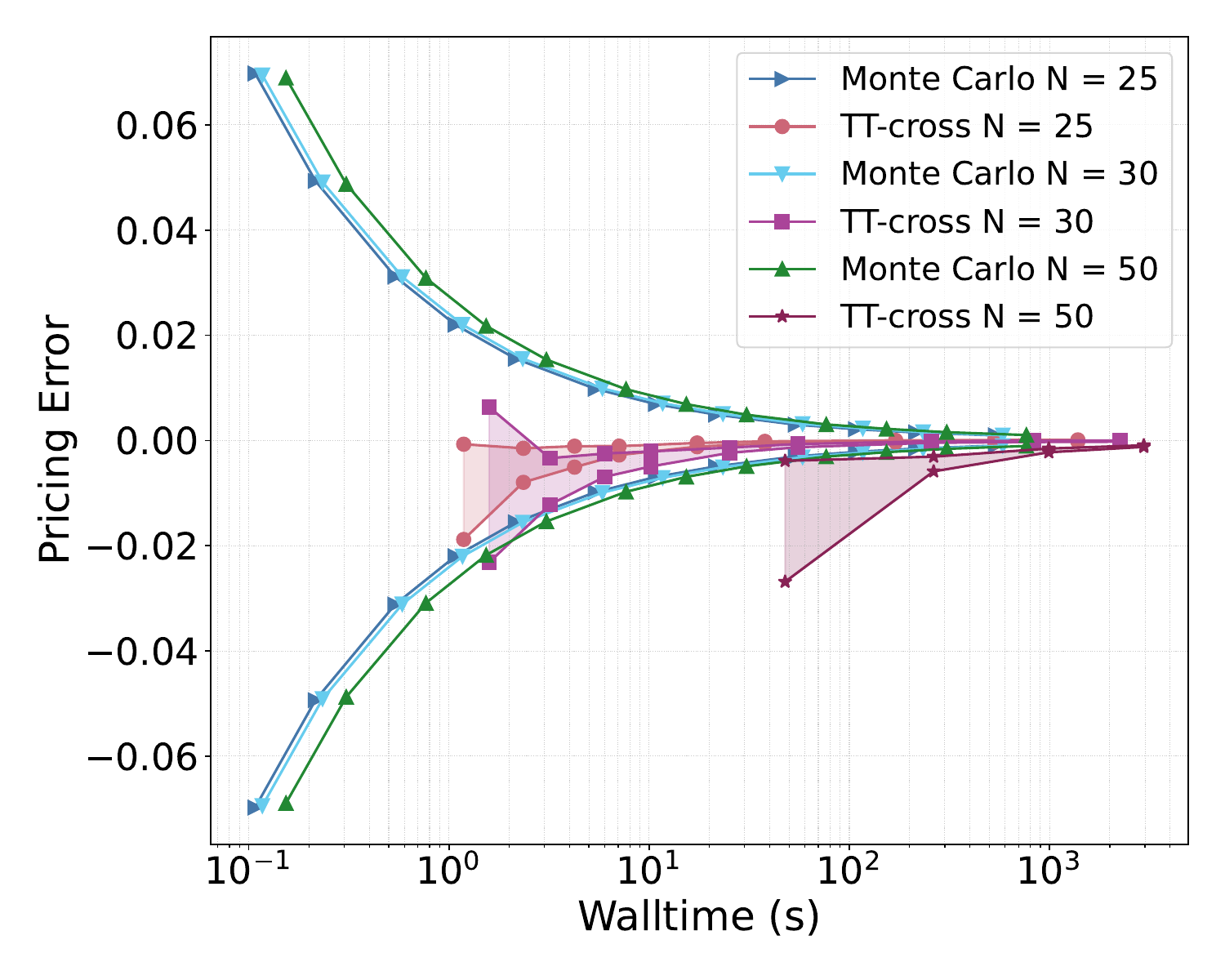}
    \caption{Comparison of the Asian option price convergence with respect to walltime between the TTcross approach and Monte Carlo sampling. Simulations are performed for a binomial model with $N = 25$, $30$, and $50$ time steps. Monte Carlo results were obtained by running across different number of samples $N_s = 10^4$ to $N_s = 5\times 10^7$ samples. For TTcross, each data point corresponds to data obtained from over $60$ independent runs performed across different bond dimensions. Finally, the median error $\pm$ one standard deviation was plotted to show convergence for both TTcross and Monte Carlo methods.}
    \label{fig:asian_ttx_vs_mc}
\end{figure}

Fig.~\ref{fig:asian_ttx_vs_mc} shows the convergence of the tensor network approach compared to conventional Monte Carlo sampling to price Asian options with the binomial pricing model for different number of time steps $N = 25$, $30$, and $50$. The $x$-axis shows the elapsed walltime in seconds for both the TTcross and Monte Carlo methods, and the $y$-axis shows the corresponding option pricing error. The Monte Carlo sampling was performed by drawing $N_s$ length-$N$ random bit strings $\mathbf{x} = x_1x_2\dots x_N$ from $p(\x)$ and computing an estimate of the average using \Eq{eq:asian-final}, i.e.
\begin{align}
    \overline{V(t=0, K|S_0)} = \frac{1}{N_s}\sum_{\x_i \sim p(\x)}^{N_s}v_A(\x_i).
\end{align}
These calculations were performed over $N_s = 10^4$ to $N_s = 5\times 10^7$ samples, with the exact options price being computed as the mean price obtained from $10^{11}$ samples. The standard deviations of the price for the different samples were then plotted on the $y$ axis, as a proxy for the pricing error, across the corresponding mean walltime incurred on the $x$ axis. 

For the TTCross method, calculations were performed across different bond dimensions ranging from $D = 30$ to $D = 250$ for $N = 25$ and $30$, and from $D = 80$ to $D = 250$ for $N = 50$. Each calculation was repeated over $60$ times. The pricing error was then computed by subtracting the exact value obtained from Monte Carlo. Finally, for each bond dimension, the spread of the error was computed by adding and subtracting the standard deviation from the median error and plotted on the $y$ axis across the corresponding mean walltime. Fig.~\ref{fig:asian_ttx_vs_mc} shows a higher convergence speed of the option price with the number of samples, and can obtain MC sampling accuracy
already at  $\approx 50$x$-100$x less walltime, especially for the smaller number of time steps.

%That is, a bit $x_i$ is $1$ with a probability $p_u$ if the asset price goes up and $0$ with a probability $1 - p_u$ if it goes down, in accordance with the binomial model. Given $\mathbf{x}$, the corresponding option price $v(\mathbf{x})$ is computed using Eq.~\ref{eq:asn_option_price_bm_traj}. Now, in classic Monte Carlo fashion, this process is repeated as many times as necessary to compute the final option price as an average of all computed values.
%The fundamental limitation of the naive Monte Carlo approach is that as the space of trajectories an asset assumes grows exponentially with the number of time steps. Hence, one cannot hope to randomly sample from an exponentially growing space to efficiently compute accurate averages. The merit of the tensor network approach lies in capturing the potentially low-dimensional nature of the set of possible prices, and then sampling only from this low-dimensional space, leading to better convergence as seen in Fig.~\ref{fig:asian_ttx_vs_mc}.

\subsubsection{Variational approach for Asian option pricing}
\label{ssec:var_asn_method}
The second method to price Asian options is based on a variational optimization approach. Consider
the function $\tilde v_A(\mathbf{x})$ obtained from dropping
the $\max$ function in \Eq{eq:asian_option_payoff},
\begin{align}
    &\tilde v_A = \langle S_T(\x)\rangle - K = \frac{S_0}{N} \sum_{i = 1}^N  \prod_{k = 1}^{i} d^{1 - x_k}u^{x_k} - K\label{eq:asian-ineq-1}\\
    &\tilde v_A(\x) \leq v_A(\x)\;\;\forall\x.\label{eq:asian-ineq-2}
\end{align}
From \Eq{eq:asian-ineq-2} it follows immediately that 
% \begin{equation}
% \begin{split}
% e^{-rT}\sum_{\mathbf{x}} p(\mathbf{x})\tilde v_A(\mathbf{x}) &\leq e^{-rT}\sum_{\mathbf{x}} p(\mathbf{x}) v_A(\mathbf{x}) \\
% e^{-rT}\sum_{\mathbf{x}} p(\mathbf{x}) v_A(\mathbf{x}) &= V(t=0,K \mid S_0)
% \end{split}
% \label{eq:asian-ineq-3}
% \end{equation}
\begin{equation}
   e^{-rT}\sum_{\x} p(\x)\tilde v_A(\x) \leq e^{-rT}\sum_{\x} p(\x) v_A(\x)=V(t=0,K|S_0)\label{eq:asian-ineq-3}
\end{equation}
with equality in \Eq{eq:asian-ineq-3} obtained when restricting the sum on the left hand side to values $\x$ for which $p(\x)\tilde v_A(\x)\geq 0$. Note that $p(\x)\tilde v_A(\x)$ has an exact, efficient MPS representation
\begin{equation}
    p(\x)\tilde v_A(\x) = B_1^{x_1} B_2^{x_2} \dots B_N^{x_N}
\end{equation}
with tensors 
\begin{align}
    B^0_i& =
    \begin{bmatrix}
        d(1-p_u) & d(1-p_u) \\
        0 & 1-p_u 
    \end{bmatrix} \; {\rm for }\; 1<i<N\\
    B^1_i&=
    \begin{bmatrix}
        up_u &up_u \\
        0 & p_u 
    \end{bmatrix}, \; {\rm for }\; 1<i<N\\
        B^0_1 &= 
        \begin{bmatrix}
            \frac{S_0}{N}d(1-p_u),& \frac{S_0}{N}d(1-p_u) - K]
        \end{bmatrix}\\
         B^1_1 &= 
         \begin{bmatrix}
           \frac{S_0}{N}up_u, &  \frac{S_0}{N}up_u - K
        \end{bmatrix}\\
        B^0_N &= 
        \begin{bmatrix}
        d(1-p_u) \\ 1-p_u 
        \end{bmatrix}\\
        B^1_N &=
        \begin{bmatrix}
        up_u\\
        p_u
        \end{bmatrix}.
\end{align}
We can formalize the idea of restricting the left-hand sum in \Eq{eq:asian-ineq-3} to values $\x$ with $p(\x)\tilde v_A(\x)\geq 0$ into
an optimization problem, and use a variational approach to solve it. Consider a binary tensor $\psi_{x_1x_2\dots x_N}$ such that
\begin{align}
    \psi_{x_1x_2 \dots x_N} \in \{0,1\}\quad  \forall \x,\label{eq:binary-tensor}
\end{align}

and the cost function $\mathcal{K}$ with
\begin{align}
    \mathcal{K} = e^{-rT}\sum_{\x}  \psi_{x_1x_2 \dots x_N} p(\x) \tilde v_A(\x).\label{eq:costfun}
\end{align}
It follows from \Eq{eq:asian-ineq-3} that maximization of the cost function $\mathcal{K}$ with respect to $\psi_{x_1x_2\dots x_N}$, yields the desired option price, i.e. 
\begin{equation}
   \mathcal{K}^*\equiv \max_{\psi_{x_1\dots x_N}}(\mathcal{K})= V(t=0,K|S_0).
\end{equation}
The optimization of \Eq{eq:costfun} is, in general, intractable due to the exponential size of $\psi_{x_1x_2\dots x_N}$. Instead, we propose to use an MPS ansatz for the binary tensor $\psi_{x_1x_2\dots x_N}$, i.e. 
\begin{equation}
    \psi_{x_1x_2\dots x_N} = A_1^{x_1} A_2^{x_2} \dots A_N^{x_N}, \label{eq:psi-mps}
\end{equation}
which implies that \Eq{eq:costfun} can be evaluated efficiently in a time linear in $N$. 

The parametrization of a binary MPS is in itself a highly non-trivial task, and we are not aware of
any previous work addressing this problem. For the purpose of this work, we use the following simple parametrization:
consider an MPS with tensors $L^{x_k}_{\alpha_{k-1}\alpha_k}$ for $k < c$, $R^{x_k}_{\alpha_{k-1}\alpha_k}$ for $k > c$ and $A^{x_c}_{\alpha_{c-1}\alpha_c}$ s.t.
\begin{align}
    \psi_{x_1\dots x_N} = L^{x_1}_1\dots L^{x_{c-1}}_{c-1}A_c^{x_c}R_{c+1}^{x_{c+1}}\dots R_N^{x_N},\label{eq:mps-LAR}
\end{align}
and with $L_{\alpha_{k-1}\alpha_k}^{x_k}, R_{\alpha_{k-1}\alpha_k}^{x_k}$ obeying the constraints
\begin{align}
    \label{eq:lprop}
    \sum_{\alpha_k} L_{\alpha_{k-1}\alpha_k}^{x_k} \in \{0, 1\}\;\forall (k<c, x_k, \alpha_{k-1})\\
    \label{eq:rprop}
    \sum_{\alpha_{k-1}} R_{\alpha_{k-1}\alpha_k}^{x_k} \in \{0, 1\}\;\forall(k>c, x_k, \alpha_{k}).
\end{align}
These conditions imply that for any choice of binary string $\x$, the arrays
\begin{align}
    [L^{x_1}_1\dots L^{x_{k}}_{k}]_{\alpha_k},\;\;k<c\nonumber\\
    [R^{x_l}_l\dots R^{x_{N}}_{N}]_{\alpha_{l-1}},\;\;l>c\nonumber
\end{align}
obtained from contracting all tensors $\{L_k^{x_k}\}$ with $k<c$ and $\{R_l^{x_l}\}$ with  $l>c$ are zero vectors or binary unit vectors with a single non-zero element (see Appendix \ref{appx:left_right_binary_property}). With this setting, it follows that any choice of binary matrix $A_c^{x_c}$ will yield a valid binary MPS representation. We emphasize that this ad-hoc choice of representation is by no means a general way of parametrizing binary MPS in \Eq{eq:psi-mps}, and for any finite bond dimension $D$ can only provide a lower bound for the actual option price. However, our numerical experiments below suggest a rapid convergence to exact results (where applicable) with increasing bond dimension, rendering the ansatz highly useful in practice.

To optimize the cost function \Eq{eq:costfun}, we use a sweeping algorithm which optimizes the MPS by updating one tensor at a time while keeping all other tensors fixed, similar to the density matrix renormalization group (DMRG) algorithm \cite{legeza_qc-dmrg_2003, white_ab_1999}. 
In the following, we describe such an update step. Starting from the decomposition \Eq{eq:mps-LAR} at $c=1$, we optimize the tensor $A_c^{x_{c}}$ while keeping all other tensors fixed. Note that for the decomposition \Eq{eq:mps-LAR} the optimal tensor $V_c^{x_c}\equiv V_{\alpha_{c-1}\alpha_c}^{x_c}$ is given by
\begin{equation}
    V_{\alpha_{c-1}\alpha_c}^{x_c} = \frac{\textrm{sign}\Bigg[\frac{\partial\mathcal{K}(L_1^{x_1}\dots L_{c-1}^{x_{c-1}} A_c^{x_c}R_{c+1}^{x_{c+1}}\dots R_N^{x_{N}})}{\partial A_{\alpha_{c-1}\alpha_{c}}^{x_c}}\Bigg] + 1}{2}
\label{eq:value_c}
\end{equation}
which can be computed using standard tensor network calculus. Next, we attempt to decompose the matrix $V_c^{x_c}$ as
\begin{equation}
    V_c^{x_c}\approx L_c^{x_c} M \label{eq:decomp}
\end{equation}
with $L_c^{x_c}$ satisfying \Eq{eq:lprop}. Note that \Eq{eq:decomp} can always be fulfilled by making the virtual dimensions large enough. We prevent the bond dimension from growing uncontrollably by constructing an approximation, for example, by minimizing:
\begin{align}
    L^{x_c}_{c}, M = \argmin_{\tilde L_c^{x_c},\tilde M}\lVert V_c^{x_c} - \tilde L_c^{x_c} \tilde M\rVert,
\end{align}
which is a mixed-integer, constrained linear optimization problem. Once this decomposition has been found, 
the matrix $m$ can be discarded, and the optimization can proceed with the next site $c + 1$.
The mixed integer optimization problem \Eq{eq:decomp} is, in general, too expensive to solve for large bond dimensions.
Instead, we construct a suboptimal solution using a greedy algorithm as described in Appendix \ref{appx:variational_greedy_algorithm}. 
Note that even though the output of the optimization in general depends on the initialization of the MPS in \Eq{eq:psi-mps}, in practice we find a rapidly diminishing influence of the initial state on the option value with increasing bond dimension $D$.

\begin{figure}
	\centering
	\begin{subfigure}{\linewidth}
		\includegraphics[width=\columnwidth]{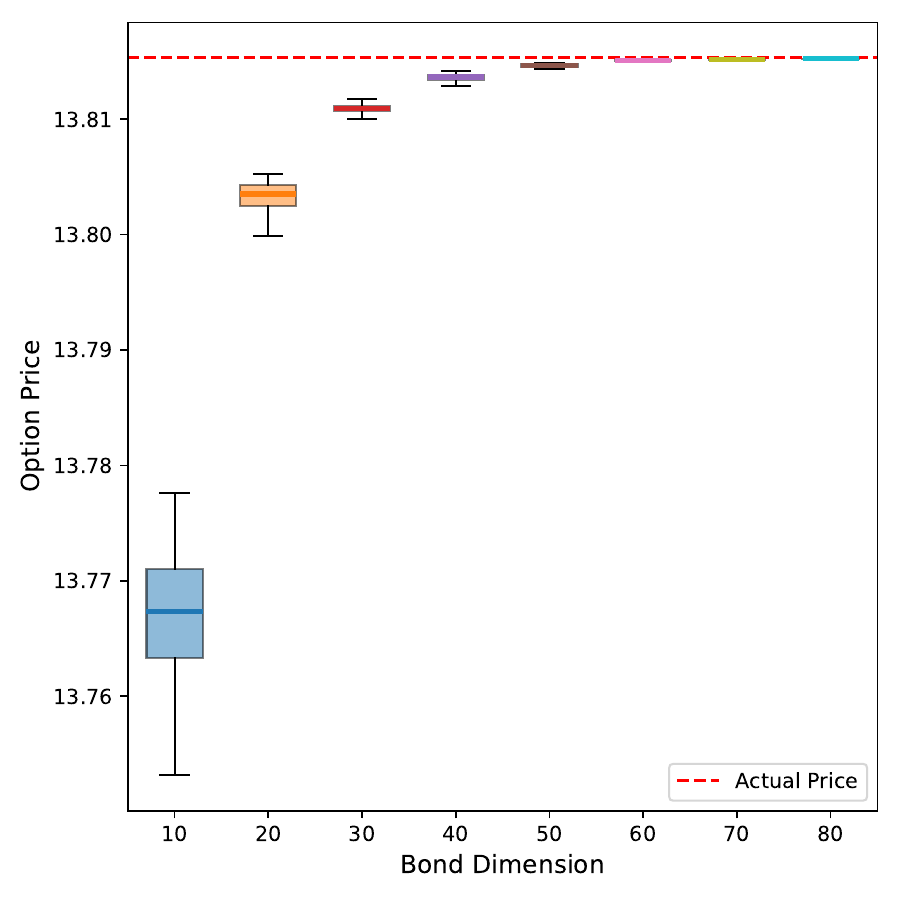}
		\caption{Number of time steps $N = 20$}
		\label{fig:var_asn_tn_price_vs_bd_N20}
	\end{subfigure}
        \\
	\begin{subfigure}{\linewidth}
		\includegraphics[width=\columnwidth]{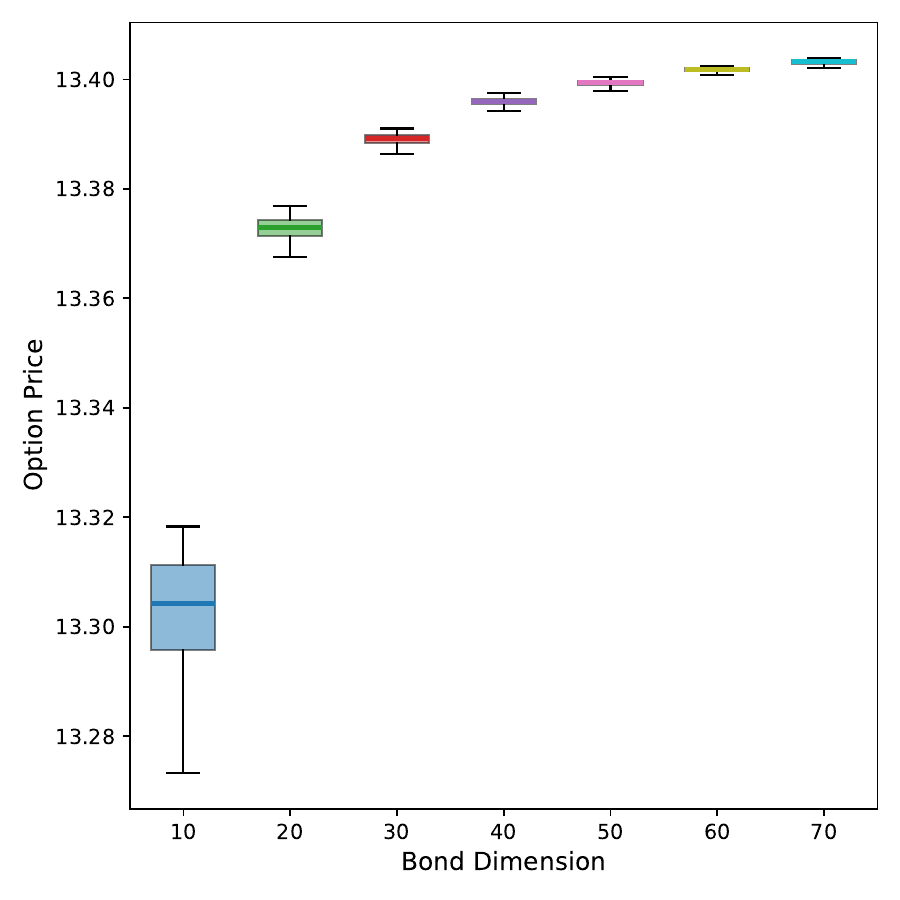}
		\caption{Number of time steps $N = 60$}
		\label{fig:var_asn_tn_price_vs_bd_N60}
	\end{subfigure}
	\caption{Convergence of the Asian option price calculated using the variational method with respect to increasing bond dimension of the binary MPS tensor as described in \Eq{eq:psi-mps}. In \Fig{fig:var_asn_tn_price_vs_bd_N20}, the exact binomial model Asian options price obtained from a brute force calculation is shown as a red dotted line for the number of time steps $N = 20$. For each bond dimension, the calculations were repeated $60$ times, and the corresponding price distributions are plotted as box plots.}
	\label{fig:var_asn_tn_price_vs_bd}
\end{figure}

In \Fig{fig:var_asn_tn_price_vs_bd} we show the convergence of the variational approach with increasing bond dimension $D$ of the MPS ansatz \Eq{eq:psi-mps}. The standard parameters described above were used to run these calculations, and for each bond dimension, the calculations were repeated $60$ times. In the case of $N = 20$ time steps, the exact price as obtained from the binomial model was calculated using brute-force, and is depicted as a red dotted line in \Fig{fig:var_asn_tn_price_vs_bd_N20}. 

\begin{figure}
    \centering
    \includegraphics[width=0.95\linewidth]{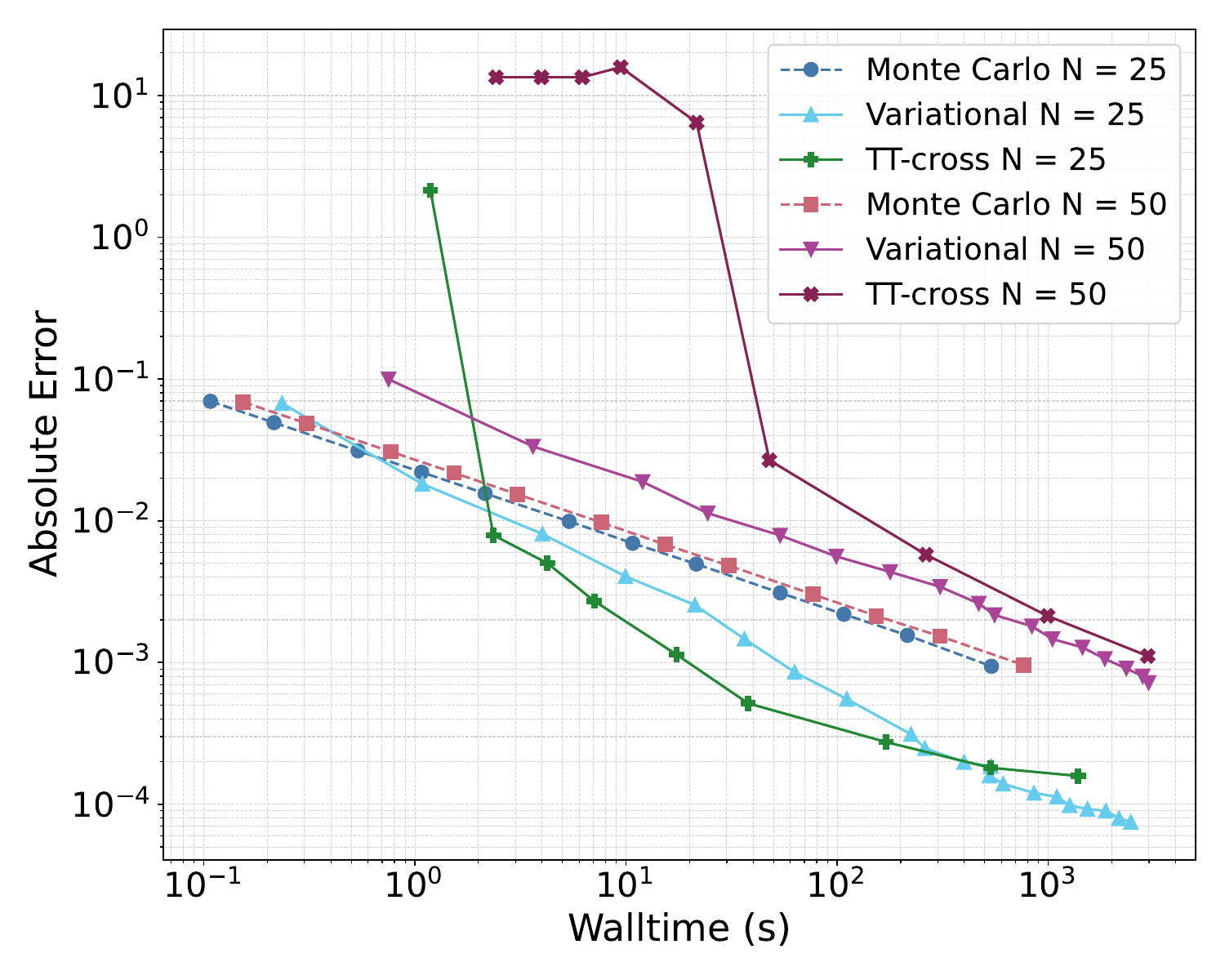}
    \caption{Pricing error comparison of the Asian option price convergence with respect to walltime for Variational, TTcross, and Monte Carlo methods for different number of time steps $N = 25$ and $N = 50$. Each data point corresponds to $60$ independent calculations.}
    \label{fig:asian_ttx_vs_mc_vs_var_diffN}
\end{figure}

\begin{figure}
    \centering
    \includegraphics[width=0.95\linewidth]{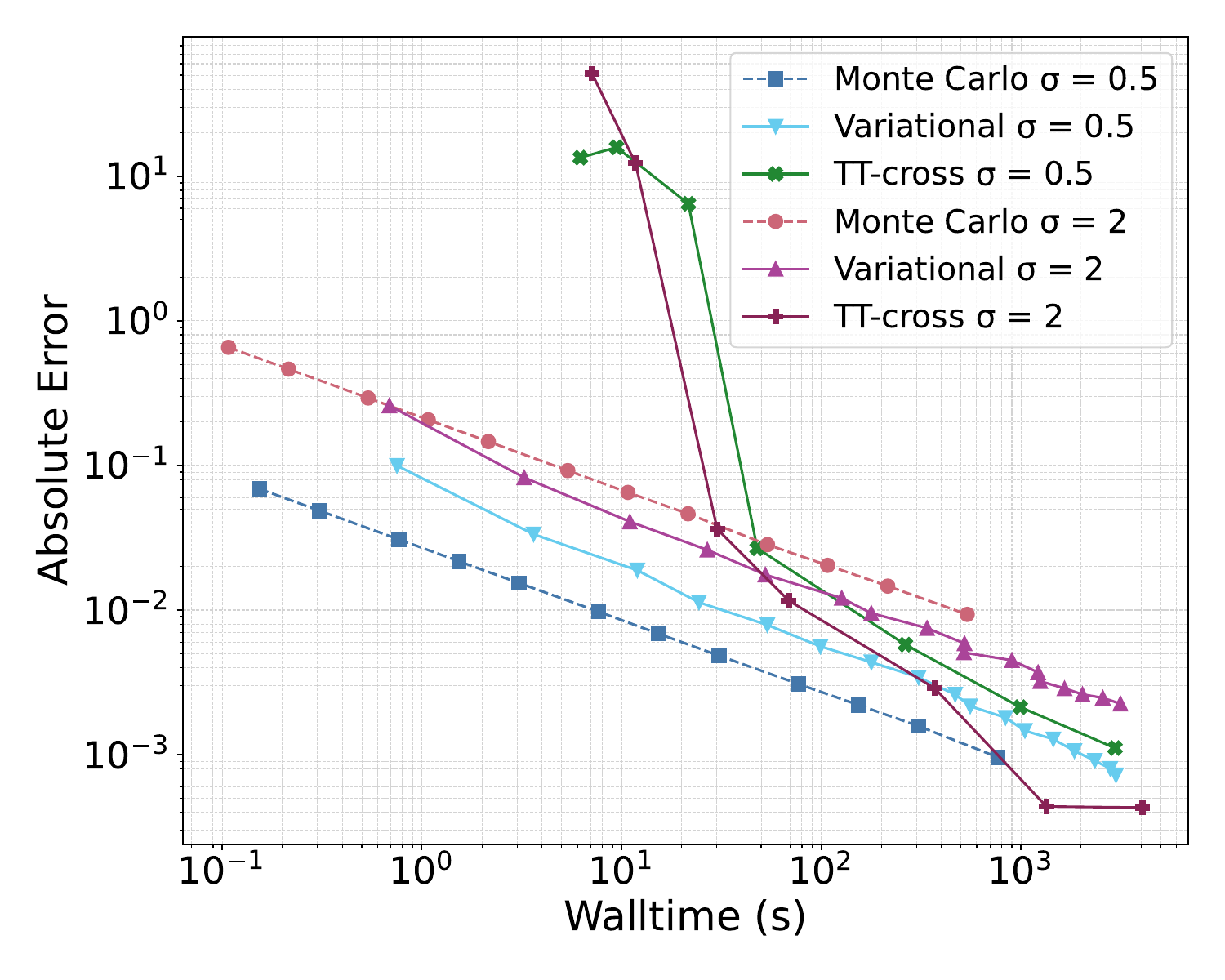}
    \caption{The role of asset volatility in pricing error convergence with respect to walltime for Variational, TTcross, and Monte Carlo methods for different asset volatilities $\sigma = 2$ and $\sigma = 0.5$. Each data point corresponds to $60$ independent calculations.}
    \label{fig:asian_ttx_vs_mc_vs_var_diffvol}
\end{figure}

In \Fig{fig:asian_ttx_vs_mc_vs_var_diffN} and \Fig{fig:asian_ttx_vs_mc_vs_var_diffvol}, we compare the performance of the three different Asian option pricing methods discussed in this paper, conventional Monte Carlo, TTcross, and variational. For each case, the exact option price was estimated through Monte Carlo simulations with $10^{11}$ samples. Next, each data point in the plot corresponds to the median error across multiple runs plus the standard deviation across those runs. For the TTcross and Variational methods, the different data points were obtained from running $60$ independent runs across different bond dimensions as in \ref{sssec:asn_ttcross_method}. 

\Fig{fig:asian_ttx_vs_mc_vs_var_diffN} compares the convergence of the three different methods for different number of time steps, $N = 25$ and $N = 50$. As expected, the Monte Carlo method is agnostic to the number of time steps, and shows a similar convergence profile for both $N$. On the other hand, both the TTcross and Variational methods show almost an order of magnitude reduction in error for the same walltime, compared to Monte Carlo for $N = 25$. However, $N = 50$ is a significantly higher-dimensional problem where we see that Monte Carlo performs better. 

The calculations in \Fig{fig:asian_ttx_vs_mc_vs_var_diffN} comparing the performance of the three methods for different number of time steps were performed for an Asian option on an asset with an underlying volatility $\sigma = 0.5$. In such a case, when the asset price does not fluctuate too widely, it is reasonable that random sampling, as in Monte Carlo, would be able to accurately describe the final option payoff distribution and hence evaluate the option's price efficiently with a smaller number of samples. This also makes sense because in the limit of zero variance, we should find the Monte Carlo method to converge instantaneously. However, in the case where the underlying asset dynamic is highly volatile, say with $\sigma = 2$, it would be significantly harder for random sampling methods like Monte Carlo to accurately price the option efficiently. And exactly this is depicted in \Fig{fig:asian_ttx_vs_mc_vs_var_diffvol}. Here, we compare the performance of the three methods for $N = 50$ across different volatilities $\sigma = 0.5$ and $\sigma = 2$. As seen above, for $\sigma = 0.5$, we see that Monte Carlo has a better convergence as compared to both TTcross and Variational methods. However, for $\sigma = 2$, we see that both TTcross and Variational methods outperform Monte Carlo, with TTcross showing more than an order of magnitude lower error rates.

% shows the convergence of the method applied to a $N=25$ and a $N=50$ period binomial model to the exact price, which can be estimated by running a large number of Monte Carlo simulations. In \Fig{fig:asian_ttx_vs_mc_vs_var_diffvol} we compare the performance of our methods in an $N=50$ period binomial method for two different volatilities. We find that the convergence speed is dependent on the number of time steps and the particular distribution.  We see that both TTcross and the variational approach can outperform Monte Carlo by orders of magnitude, but not across the board.

\subsection{American basket option pricing with matrix product states}
The binomial pricing algorithm for standard single-asset options can be straightforwardly generalized to the
multi-asset case \Eq{eq:gbm_multi-asset}, but faces an exponentially increasing cost \cite{Mller2009TheBA} with the number of underlying assets $m$ (see Appendix~\ref{appx:binomial_pricing_of_amr_options}). In this work, we focus on the decoupled trees approach for binomial multi-asset option pricing, which is an improved version of the standard binomial method for multi-asset options \cite{Korn2009p1, Korn2009p2, Mller2009TheBA}. In the following, we will briefly outline the basic ideas and refer the reader to \cite{Mller2009TheBA} for a detailed introduction.

Given a basket with $m$ correlated assets $S^i_t$ described by \Eq{eq:gbm_multi-asset}, we use Ito's lemma \cite{ito, hull2021, Jacobs_2010} to apply a logarithmic transformation to obtain the variables $X^i_t$, i.e. 
\begin{align}
\label{eq:log_transform_gbm}
    X^i_t &\equiv \ln S^i_t\\
    dX^i_t &= (r - \frac{1}{2} \sigma^2_i)dt + \sigma_i dW^i_t.
\end{align}
The basic idea of the decoupling approach is to transform the correlated variables $X^i_t$ to a set of uncorrelated variables $Y^i_t$ with a diagonal correlation matrix. One way to achieve this is through a Cholesky decomposition of the covariance matrix $\Sigma_{ij}$ in \Eq{eq:covar}, i.e. 
\begin{equation}
    \begin{split}
        \Sigma &= GG^T, \\
        \mathbf{Y}_t &= G^{-1}\mathbf{X}_t\\
    \end{split}
\end{equation}
with lower triangular matrix $G$, such that
\begin{align}
        dY^j_t = \alpha_j dt + d\bar{W}^j_t\nonumber\\
        \pmb{\alpha} = G^{-1}\left( r\mathbbm{1} - \frac{1}{2}\pmb{\sigma}^2 \right),
\end{align}
where use bold letters to denote array quantities, i.e. $\mathbf{X}_t = (X^1_t, \dots, X^m_t)$ and similar for $\mathbf{Y}_t, \pmb{\alpha}$ and $\pmb{\sigma}$. $d\bar W^j_t$ here are $m$ uncorrelated random walks with $\mathbb{E}[d\bar W^i_td\bar W^j_t]=\delta_{ij} dt$.
In this work, we choose a Cholesky decomposition of the correlation matrix, but other choices have been explored
in the literature \cite{Mller2009TheBA}. The continuous evolution of the $m$ uncorrelated variables $\mathbf{Y}_t$
can then be approximated by $m$ independent binomial trees \cite{Mller2009TheBA} with variables $Y^i_{t_k}, k = \{0,1,\dots N\}, i = \{1,2,\dots,m\}$.
In the following, we use the Rendleman-Bartter model to discretize each independent random walk, with up-move and down-move factors $u_i$ and $d_i$ given by
\begin{equation}
    \begin{split}
        u_i &= \alpha_i \Delta t + \sqrt{\Delta t}, \\
        d_i &= \alpha_i \Delta t - \sqrt{\Delta t}
    \end{split}
\end{equation}
and up- and down-move probabilities $(p_u^i, p_d^i) = (\frac{1}{2}, \frac{1}{2})\;\forall i$.
At any discrete time step $k$ of the multi-dimensional tree, the original random variables $\mathbf{S}_{t_k}$ are given by
\begin{equation}
    \mathbf{S}_{t_k}= \exp(G \mathbf{Y}_{t_k}).
\end{equation}
To reduce the burden on index notation, we will in the following use the abbreviation $\mathbf{Y}_k\equiv \mathbf{Y}_{t_k}$.

The key quantity in the decoupling approach to multi-asset binomial option pricing is the payoff function at expiration $t_N=T$ of the option. For a Put option with max-basket payoff function, this is given by
\begin{equation}
\label{eq:amr_basket_expr_payoff}
    \begin{split}
        v(\mathbf{Y}_N) &= \text{max}\left( K - \text{max}\left( e^{G\mathbf{Y}_N} \right), 0 \right).
    \end{split}
\end{equation}
%where $\mathbf{Y}_N$ are uncorrelated, discrete random variables at discrete expiration time $t_N = T$ (see section \ref{sec:binomial_model}). 
Representing this function on $m$ random variables in general scales exponentially in $m$ for generic payoff functions. 
In this work, we use the tensor train cross approximation to learn an approximation of this function in a
time linear in $m$, i.e.
\begin{equation}
\label{eq:amr_basket_ttcross}
    \begin{split}
        v(\mathbf{Y}_N) \approx M_1^{[N],Y_N^1} M_2^{[N],Y_N^2} \dots M_m^{[N],Y_N^m}.
    \end{split}
\end{equation}
where $M^{[N],Y^i_N}_{i}$ are the matrices of the matrix product state at step $N$ of the binary tree.
We slightly abuse notation here by assuming variables $Y^i_N\in (0,1,\dots,N)$ to take integer values, corresponding to the different possible outcomes
\begin{align}
 (Y_0^id^N, Y_0^id^{N-1}u, Y_0^id^{N-2} u^2, \dots, Y_0^id u^{N-1}, Y_0^iu^N)\label{eq:rvoutcomes}
\end{align}
at $t_N=T$, with $Y_0^i = [G^{-1} \mathbf{X}_{t=0}]_i$. 

{\bf European basket option:} In the European case, exercising is only possible at expiration. 
The option value is, in this case, given by
\begin{align}
    V(&t=0,K|\mathbf{S}_0) = \\
    &=e^{-rT}\sum_{\mathbf{Y}_N}v(\mathbf{Y}_N)p(\mathbf{Y}_N)\nonumber\\
    &=e^{-rT}\sum_{\mathbf{Y}_N, \mathbf{Y}_{N-1}}v(\mathbf{Y}_N)p(\mathbf{Y}_N|\mathbf{Y}_{N-1})p(\mathbf{Y}_{N-1}) \nonumber\\
    \vdots\nonumber\\
    &=e^{-rT}\sum_{\mathbf{Y}_N\dots \mathbf{Y}_{1}}v(\mathbf{Y}_N)p(\mathbf{Y}_N|\mathbf{Y}_{N-1})\dots p(\mathbf{Y}_1|\mathbf{Y}_0), \label{eq:european-multi-expanded}
\end{align}
with $p(\mathbf{Y}_N) = p(Y_1) \dots p(Y_N)$ the probability of observing the $m$ independent outcomes $\mathbf{Y}_N$ of the decoupled binomial tree.
With the definition of the conditional expectation value $\mathbb{E}[\
\cdot |\cdot ]$ for a function $\mathcal{V}^k(\mathbf{Y}_k)$, 
\begin{align}
    \mathbb{E}[\mathcal{V}^k(\mathbf{Y}_{k})|\mathbf{Y}_{k-1}]&\equiv\sum_{\mathbf{Y}_{k}} p(\mathbf{Y}_{k}|\mathbf{Y}_{k-1})\mathcal{V}^k(\mathbf{Y}_k)
\end{align}
the expectation value \Eq{eq:european-multi-expanded} can be computed recursively from
\begin{align}
    \mathcal{V}^N(\mathbf{Y}_N) &= v(\mathbf{Y}_N) \nonumber\\
    \mathcal{V}^{k-1}(\mathbf{Y}_{k-1})&= \mathbb{E}[\mathcal{V}^{k}(\mathbf{Y}_{k})|\mathbf{Y}_{k-1}]\label{eq:recurr}\\
    V(t=0,K|\mathbf{S}_0) &= e^{-rT} \mathcal{V}^0(\mathbf{Y}_0),\nonumber
\end{align}
i.e. \Eq{eq:european-multi-expanded} reduces to
\begin{align}
    V(&t=0,K|\mathbf{S}_0) =\mathbb{E}[\mathbb{E}[\dots \mathbb{E}[v(\mathbf{Y}_{N})|\mathbf{Y}_{N-1}] \dots]|\mathbf{Y}_0].\label{eq:recurr-2}
\end{align}
Given the MPS representation \Eq{eq:amr_basket_ttcross} of $\mathcal{V}^N(\mathbf{Y}_N) = v(\mathbf{Y}_N)$, the conditional expectation at step $N-1$  is given by
\begin{align}
    &\mathcal{V}(\mathbf{Y}_{N-1}) =\sum_{\mathbf{Y}_N}p(\mathbf{Y}_N|\mathbf{Y}_{N-1})M_1^{[N],Y_N^1} M_2^{[N],Y_N^2} \dots M_m^{[N],Y_N^m}\nonumber\\
    &=\sum_{\mathbf{Y}_N}p(Y_N^1|Y_{N-1}^1)\dots p(Y_N^m|Y_{N-1}^m) M_1^{[N],Y_N^1}\dots M_m^{[N],Y_N^m}\label{eq:mps-european-multi}
\end{align}
and can be computed efficiently using standard tensor network techniques and the matrix form of conditional 
probabilities $p(Y^i_N|Y^i_{N-1})$, see \Eq{eq:probmatrix} in the Appendix~\ref{appx:basket_probability_operators}. Note that given the MPS representation of $\mathcal{V}(\mathbf{Y}_{N})$, $\mathcal{V}(\mathbf{Y}_{N-1})$ is in this case again in an MPS format with matrices
\begin{align}
    M^{[N-1], Y^i_{N-1}}_{i} =\sum_{Y_N^i=0}^N p(Y^i_{N}|Y^i_{N-1}) M^{[N], Y^{i}_{N}}_{i}.
\end{align}
The final option price $V(t=0, K|\mathbf{S}_0)$ is then obtained from repeated application of \Eq{eq:recurr} (see also the Appendix for an alternative derivation).

{\bf American basket option:} Unlike the European case, we now need to take into account the possibility of exercising at any time $t_k \leq t_N$. Starting again from an MPS approximation to the payoff $v(\mathbf{Y}_k)$ at expiry $k=N$, and taking into account the possibility of early exercise, the conditional expectation value of the payoff at the previous step $k-1$ at time $t_{k-1}$ is now given
\begin{align}
    \mathcal{V}^{k-1}(\mathbf{Y}_{k-1}) = \max[e^{-r\Delta t}\mathbb{E}[&\mathcal{V}^k(\mathbf{Y}_{k})|\mathbf{Y}_{k-1}],v(\mathbf{Y}_{k-1})].\label{eq:american-basket-payoff}
\end{align}
with $\Delta t = t_{k} - t_{k-1}$. Note that while the computation of the conditional expectation $\mathbb{E}[\mathcal{V}^k(\mathbf{Y}_{k})|\mathbf{Y}_{k-1}]$ can be carried out
as before in the European case, $\mathcal{V}^{k-1}(\mathbf{Y}_{k-1})$ is now in general no longer in an MPS representation, even for the case of $\mathcal{V}^k(\mathbf{Y}_k)$ being in MPS format.
To recover the recursive evaluation scheme of \Eq{eq:recurr-2}, we use the TTcross method to obtain an MPS approximation of \Eq{eq:american-basket-payoff} at every step $k$, see Algorithm \ref{alg:mps_american_put_basket} for details. With this additional approximation, the approximate option value $V(t=0,K|\mathbf{S}_0)$ can be obtained in a time linear in the number of assets $m$.

\begin{algorithm}[H]
\caption{Pricing American basket put options using MPS and decoupled binomial trees}
\begin{algorithmic}[1]
\Require{$\mathbf{S}_0$: current stock price of the $m$ assets, $K$: strike price, $r$: risk-free rate, $\mathbf{\sigma}$: volatilities of $m$ assets, $\Sigma$: asset value correlation matrix,  $T$: time to expiration, $N$: number of time steps}
\State $\Delta t \gets \frac{T}{N}$\Comment{time step}
\State $G \mid \Sigma = GG^T$\Comment{Cholesky decomposition}
\State $\mathbf{X}_0 \gets \ln(\mathbf{S}_0)$\Comment{Log-transformation}
\State $\mathbf{Y}_0 \gets G^{-1}\mathbf{X}_0$\Comment{Uncorrelated random variables}
\State $\pmb{\alpha} \gets G^{-1}\left( r\mathbbm{1} - \frac{1}{2}\pmb{\sigma}^2 \right)$\Comment{modified drift vector}
\State $u_i \gets \alpha_i \Delta t + \sqrt{\Delta t}$\Comment{modified RB up-factor}
\State $d_i \gets \alpha_i \Delta t - \sqrt{\Delta t}$\Comment{modified RB down-factor}
\State $p_u \gets \frac{1}{2}$\Comment{RB model up movement probability}

\State $v(\mathbf{Y}_N) \gets M_1^{[N],Y_N} M_2^{[N],Y_N^1} \dots M_m^{[N],Y_N^m}.$\Comment{Approximate the option values at expiration as an MPS using TTcross}
\For{$k \gets N$ to $1$}\Comment{Backward induction}
    \State $W_i^{Y_{k-1}^{i}}\gets \sum_{Y_k^i}p(Y^i_{k}|Y^i_{k-1})M_i^{[k],Y_k^{i}} \ \forall \ i \in \{1, \dots, m\}$
    \State $\mathcal{V}^{k-1}(\mathbf{Y}_{k-1}) \gets M_1^{[k-1],Y_{k-1}^1} M_2^{[k-1],Y_{k-1}^2} \dots M_m^{[k-1],Y_{k-1}^m    }$    \Comment{Approximate$\mathcal{V}^{k-1}(\mathbf{Y}_{k-1}) $ as an MPS using TTcross}
\EndFor
\State \Return $\Pi_{i = 1}^m M_i^{[0],Y_0^i}$\Comment{Option price}
\end{algorithmic}
\label{alg:mps_american_put_basket}
\end{algorithm}

\begin{figure}
	\centering
	\begin{subfigure}{0.95\linewidth}
		\includegraphics[width=\linewidth]{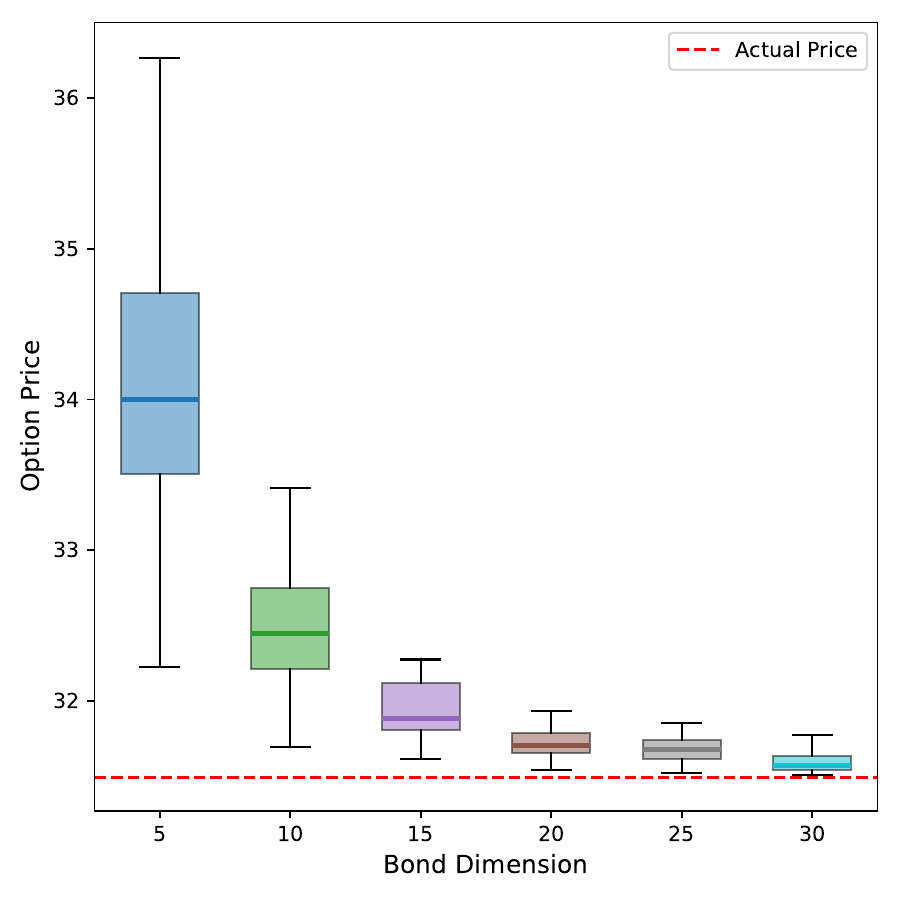}
		\caption{$m$ = 4 assets}
		\label{fig:amr_bask_tn_bm_4}
	\end{subfigure}
        \\
	\begin{subfigure}{0.95\linewidth}
		\includegraphics[width=\linewidth]{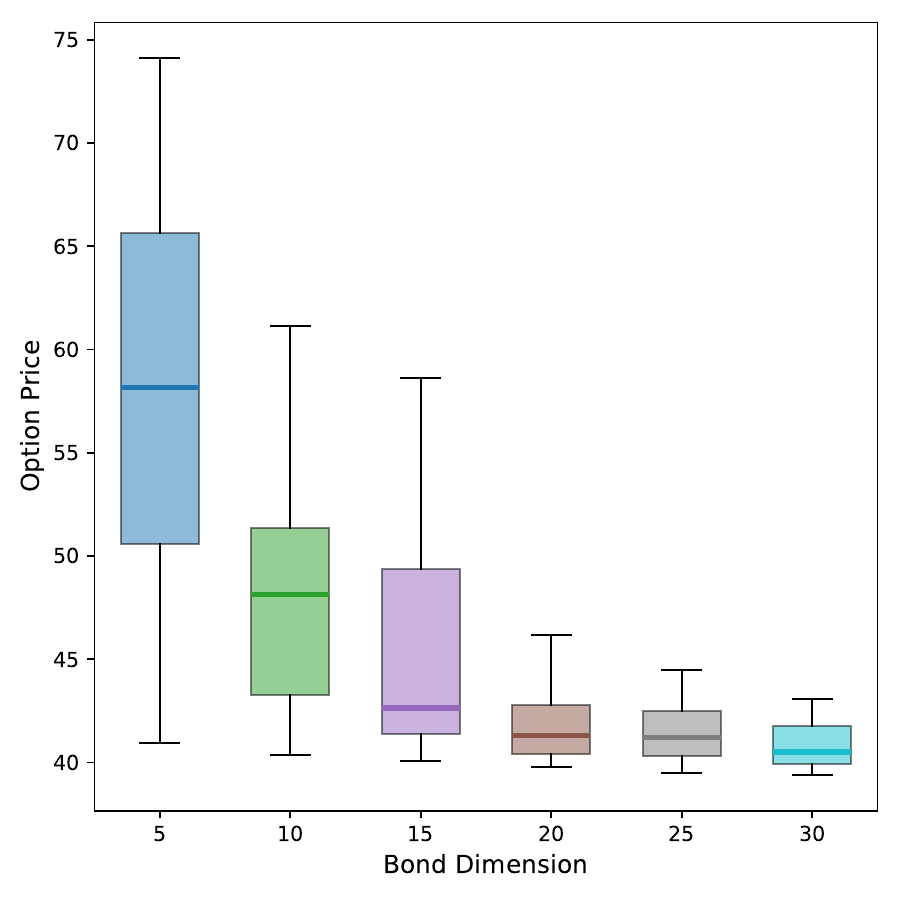}
		\caption{$m$ = 8 assets}
		\label{fig:amr_bask_tn_bm_8}
	\end{subfigure}
	\caption{Convergence of American Min Put Basket option price with increasing bond dimension obtained from using TTcross on an $N = 40$ period binomial model for $m = 4$ and $m = 8$ asset baskets. For $m = 4$ assets, the actual price obtained from brute force calculations on the binomial model is shown as a dotted horizontal line.}
	\label{fig:amr_bask_tn_price_vs_bd}
\end{figure}

The performance of the TTcross method applied to price American basket options was evaluated numerically and presented in Fig.~\ref{fig:amr_bask_tn_price_vs_bd}. For these numerical experiments, we use the same volatilities and initial asset values for all assets in the basket such that $\sigma_i = 0.5$, and $S^i_0 = 100 \ \forall \ i \in \{1, \dots, m\}$. Finally, to capture the multi-asset correlation, we use a correlation matrix $\Sigma$ where all diagonal entries are set to $1$ and non-diagonal entries are set to $\frac{1}{3}$. This formalism, although artificial, enables us to capture the correlation between assets using a single parameter. Additionally, this form also ensures that the correlation matrix is positive definite as required by the decoupling trees approach. A common risk-free interest rate of $r = 0.1$ was also used.

Fig.~\ref{fig:amr_bask_tn_price_vs_bd} shows the convergence of the option price with respect to the bond dimension for basket options with $m = 4$ (Fig.~\ref{fig:amr_bask_tn_bm_4}), and $m = 8$ (Fig.~\ref{fig:amr_bask_tn_bm_8}) assets, on using a binomial model with $N = 40$ time steps. For each bond dimension, the calculations were repeated $25$ times with the data presented as box plots. The dashed red line in \Fig{fig:amr_bask_tn_bm_4} shows the exact price for the $m = 4$ basket option computed from a brute force approach over all $40^4$ possible asset trajectories. We observe rapid monotonic convergence of the option price already at small values of the MPS bond dimensions, despite the multi-tier approximations performed to account for early exercise.

\section{Conclusions}
In this work, we investigate the application of the binomial pricing method to exotic options, specifically Asian options and American basket options, for which standard binomial approaches scale exponentially in the number of discretization steps or basket size, respectively. 
We propose the use of tensor network based approximation methods to overcome this exponential scaling. For Asian and American basket options, we employ the tensor-train cross approximation \cite{DOLGOV2020, OSELEDETS2010} to compute an approximate solution of the binomial pricing approach. Furthermore, for Asian options, we additionally propose a matrix product state based variational pricing approach, which constructs an approximate solution to the pricing problem. All approaches can be systematically improved by increasing the bond dimension of the employed tensor networks, and scale linearly in the number of discretization steps or basket sizes.

Our work demonstrates that tensor network methods—specifically Matrix Product States (MPS) and the tensor train cross (TTcross) approximation—can overcome the exponential scaling inherent in traditional binomial pricing models for exotic options. For Asian options with a lower number of time steps, the TTcross approach achieves pricing accuracy comparable to Monte Carlo methods while using roughly $50–100$ times less walltime, and under high volatility $(\sigma=2)$ both the TTcross and variational MPS methods reduce the pricing error by more than order of magnitude compared to Monte Carlo. Crucially, the variational MPS approach not only approximates the option price efficiently but also provides a rigorous lower bound—a clear advantage over Monte Carlo estimates, which can fluctuate above or below the true value.

Furthermore, the novel parametrization of a binary tensor as an MPS, along with the greedy algorithm employed for compressing and optimizing the MPS representation, shows significant promise beyond option pricing. We anticipate that this innovative framework can be applied to other combinatorial optimization problems where discrete, constrained decision variables are involved. But on the other hand, our methods do exhibit regime-specific performance, demonstrating particular benefits for smaller numbers of time steps and high volatility.  It also inherits limitations from the underlying decoupled trees approach to binomial models for multi-asset American options. However, the potential for extending these techniques to other exotic payoffs (such as barrier or look-back options) and alternative asset dynamics is clear. Future research should focus on refining these tensor network approaches and exploring their application to a broader range of financial and combinatorial optimization problems, paving the way for interdisciplinary advances in both computational finance and optimization theory.

\clearpage
\appendix

\section{Explanation on the binary nature of L and R tensors}
\label{appx:left_right_binary_property}

As stated in \Eq{eq:mps-LAR},

\begin{align*}
    \psi_{x_1\dots x_N} = L^{x_1}_1\dots L^{x_{c-1}}_{c-1}A_c^{x_c}R_{c+1}^{x_{c+1}}\dots R_N^{x_N},
\end{align*}
the tensors $L_{\alpha_{k-1}\alpha_k}^{x_k}, R_{\alpha_{k-1}\alpha_k}^{x_k}$ obey the constraints
\begin{align*}
    \sum_{\alpha_k} L_{\alpha_{k-1}\alpha_k}^{x_k} \in \{0, 1\}\;\forall (k<c, x_k, \alpha_{k-1})\\
    \sum_{\alpha_{k-1}} R_{\alpha_{k-1}\alpha_k}^{x_k} \in \{0, 1\}\;\forall(k>c, x_k, \alpha_{k}).
\end{align*}

In this section, we will describe how this leads to the resulting contracted tensors 
\begin{align*}
    [L^{x_1}_1\dots L^{x_{k}}_{k}]_{\alpha_k},\;\;k<c\nonumber\\
    [R^{x_l}_l\dots R^{x_{N}}_{N}]_{\alpha_{l-1}},\;\;l>c\nonumber,
\end{align*}
obtained from contracting all tensors $\{L_k^{x_k}\}$ with $k<c$ and $\{R_l^{x_l}\}$ with  $l>c$, are zero vectors or binary unit vectors with a single non-zero element.

For simplicity, we assume a fixed bond dimension $D$ across all indices $\alpha_{k}$. Then, for a fixed $x_k$, all \Eq{eq:lprop} is telling us is that $L^{x_k}$ are $D \times D$ binary matrices where each row has at most a single non-zero element. Similarly, \Eq{eq:rprop} states that $R^{x_k}$ are $D \times D$ binary matrices where each column has at most a single non-zero element. Next, given two such binary matrices $L^{x_k}$ and $L^{x_{k+1}}$, it suffices to show that their product is also another binary matrix that satisfies the same property. 

From elementary linear algebra, we know that left multiplication of a matrix with a row vector is equivalent to taking the linear combination of the rows of that matrix with the coefficients from the row vector. Now consider a row of the matrix $L^{x_k}$ where the $\alpha_{k_i}^{th}$ element is $1$. Left multiplying this row with $L^{x_{k+1}}$ hence corresponds copying the $\alpha_{k_i}^{th}$ row of $L^{x_{k+1}}$. Given that each row of $L^{x_{k+1}}$ also satisfies the binary property, and expanding to all rows of $L^{x_{k}}$, this means that their product also satisfies the binary property. Finally, given the first tensor $L^{x_{1}}$ is a vector, the entire product turns into a zero or a binary unit vector. The same argument applies to the $R^{x_k}$ tensors too. Finally, under this setting, the central tensor $A_c^{x_c}$ as described in \Eq{eq:mps-LAR} can take the form of any arbitrary binary tensor as it is sandwiched on both sides by zero or binary unit vectors.

\section{Greedy Algorithm for Variational Asian Options Pricing}
\label{appx:variational_greedy_algorithm}

The objective is to decompose the binary filter MPS, denoted by $V$, into a product of a binary matrix $L$ and an auxiliary matrix $M$ such that
\begin{equation}
A = L \cdot M,
\end{equation}
where $A$ is the matrix representation of the optimal local tensor derivative 
\begin{equation}
A = \frac{\partial \mathcal{K}}{\partial A},
\end{equation}
obtained by contracting the exact payoff MPS with the environments.

If one were to choose $L$ as an identity matrix of the appropriate size, the equation would be exactly satisfied. However, this choice would cause the dimension of $L$ to double after each site, leading to an exponential growth in the bond dimension. To prevent this, the algorithm employs a greedy compression procedure that iteratively reduces the number of columns (or active degrees of freedom) in $L$.

In this greedy algorithm, each row of $A$ is assigned a \emph{row positive sum} defined by
\begin{equation}
\text{rowvals}[i] = \sum_j \max\big(A[i,j], 0\big),
\end{equation}
which measures the positive contribution of row $i$. Additionally, for every pair of rows $i$ and $j$, a \emph{merge benefit} is computed as
\begin{equation}
\begin{split}
\text{mergevals}[i,j] &= \sum \max\big(A[i,:] + A[j,:], 0\big) \\
                      &\quad - \text{rowvals}[i] - \text{rowvals}[j].
\end{split}
\end{equation}
This merge benefit quantifies whether merging rows $i$ and $j$ yields a net gain in the positive contributions relative to keeping them separate.

The algorithm proceeds iteratively as follows. At each iteration, it identifies the row with the smallest positive contribution (i.e., the minimum entry in $\text{rowvals}$) and also finds the pair of rows that offers the maximum merge benefit. If the gain from merging (i.e., the merge benefit) outweighs the loss incurred by dropping the minimally contributing row, the algorithm merges the corresponding rows. Merging is performed by adding the entries of one row to the other and combining their corresponding binary filter entries using a logical OR. If merging is not beneficial, the row with the minimal contribution is dropped. After each merge or drop, the active dimension of $L$ is updated, and the relevant row and merge values are recalculated.

This greedy compression continues until the number of columns in $L$ is reduced to the target size (which is equal to the number of columns of $A$), thereby keeping the bond dimension under control while still approximating the optimal binary update for the filter MPS.

\section{Binomial pricing of American Options}
\label{appx:binomial_pricing_of_amr_options}

\begin{algorithm}[H]
\caption{American put options pricing using the CRR binomial model}
\begin{algorithmic}[1]
\Require{$S_0$: current stock price, $K$: strike price, $r$: risk-free rate, $\sigma$: volatility, $T$: time to expiration, $N$: number of time steps}
\State $\Delta t \gets T/N$: time step
\State $u \gets e^{\sigma\sqrt{\Delta t}}$: up-factor
\State $d \gets 1/u$: down-factor
\State $p \gets \frac{e^{r\Delta t} - d}{u - d}$: risk-neutral up movement probability

\State \textbf{Build stock price tree:}
\For{$i \gets 0$ to $N$}
    \For{$j \gets 0$ to $i$}
        \State $S_{i,j} \gets S_0 \cdot u^j \cdot d^{i-j}$
    \EndFor
\EndFor

\State \textbf{Compute option values at expiration:}
\For{$j \gets 0$ to $N$}
    \State \textbf{or} $V_{N,j} \gets \max(K - S_{N,j}, 0)$
\EndFor

\State \textbf{Backward induction:}
\For{$i \gets N-1$ to $0$}
    \For{$j \gets 0$ to $i$}
        \State $H_{i,j} \gets e^{-r\Delta t} \left(p \cdot V_{i+1,j+1} + (1-p) \cdot V_{i+1,j}\right)$
        \State \textbf{or} $E_{i,j} \gets \max(K - S_{i,j}, 0)$
        \State $V_{i,j} \gets \max(H_{i,j}, E_{i,j})$
    \EndFor
\EndFor

\State \textbf{Option price:}
\State \Return $V_{0,0}$
\end{algorithmic}
\label{alg:vanilla_american_put_bm}
\end{algorithm}

In this section, we shall go over how to price vanilla American put options using the binomial model with a $3$-period example. As it is well known that the optimal exercise policy for an American call option is to hold till expiration, in this paper, we will be focusing on pricing American put options. Step one is to evaluate the binomial model parameters such as the up and down factors $u$ and $d$, and the risk-neutral probability $p$ of the asset price going up, depending on the particular binomial model that is chosen. Next, we create the asset price binomial tree as shown in Fig.~\ref{fig:3pBinomialTree} where the asset value at each node $S_{i,j} = S_0 \cdot u^j \cdot d^{i-j}$. At expiration, the only choice is to exercise the option or not, depending on whether the payoff is non-zero or not. To this end, compute the payoff $\text{max}(K - S_3, 0)$ at each of the leaves in the binomial tree. Next, we use the valuations of the option at time $t = 3$ to evaluate the value at the previous time step. Given this is an American option with the possibility of early exercise, the two choices available to the contract holder are to exercise the option now, given the asset value at the present, or to hold the contract till the next time step. This decision is made by comparing the payoff obtained via exercise and the discounted expected payoff under risk-neutral probabilities received on holding the asset. The value of the option at that point would be the maximum value among these two possibilities. The same process is repeated by moving backwards through the tree till one reaches $t = 0$, at which point, the remaining value is the no-arbitrage price of the option. Alg.~\ref{alg:vanilla_american_put_bm} goes over the general procedure of pricing vanilla American options using the CRR binomial model in more detail.

\section{more on binomial pricing of multi-asset European options}
\label{appx:basket_probability_operators}

The value of the option at the current time $t_0=0$ is in this case given by the discounted expectation value of the payoff function $v(\mathbf{Y}_N)$ at $t_N=T$ taken over all possible paths $\mathbf{Y}_0, \mathbf{Y}_1, \dots ,\mathbf{Y}_N$. For an RB binomial tree of uncorrelated random variables $Y^i$, each path is equally likely. The sum over all paths
can in this case be replaced by a weighted sum over all outcomes $Y^i_N$ at termination, with the probability
\begin{equation}
    p(Y^i_N) = \#_{Y^i_N} \times \frac{1}{2^n}.
\end{equation}
where $ \#_{Y^i_N}$ counts the number of paths ending at value $Y^i_N$, which can be straightforwardly computed (see below). 
Given an MPS approximation \Eq{eq:amr_basket_ttcross} of the payoff $v(\mathbf{Y}_N)$, the option value at $t_0=0$ is then simply given by
\begin{align}
    V(&t=0,K|S_0) = \\
    &=e^{-rT}\sum_{\{Y^1_N, \dots Y^m_N\}}p(Y^1_N) \dots p(Y^m_N) M_1^{[N],Y_N^1} \dots M_m^{[N],Y_N^m}
\end{align}
which can be computed efficiently using standard tensor network techniques. The total cost of the pricing method scales linearly in the number of assets $m$.

Consider now a single binomial tree with random variable $Y_k, k=0\dots n$. At each $k$, the variable $Y_k$ can either move up or down by factor $u$ or $d$, respectively, i.e. $Y_{k+1} = uY_k$ or $Y_{k+1} = dY_k$.
For convenience, we will label outcomes of the random variable $Y_k$ with integer values, where each distinct integer corresponds to a possible observed outcome, i.e. 
\begin{align}
  Y_k &= 0 \;\;\hat = \;Y_0d^k\nonumber\\
  Y_k &= 1 \;\;\hat = \;Y_0d^{k-1}u\nonumber\\
  Y_k &= 2 \;\;\hat = \;Y_0d^{k-2} u^2\nonumber\\
  &\vdots\\
  Y_k &= k \;\;\hat = \; Y_0u^k.\label{eq:probmatrix}
\end{align}
We are interested in computing the number of paths leading to a final observation $Y_N=y_N \in \{0,1,\dots N\}$ at $k=N$. To this end we introduce the matrices $P_{Y_kY_{k-1}}$ with 
\begin{align}
    P_{Y_0} = 
    \begin{bmatrix}
1
\end{bmatrix}\\
P_{Y_1Y_0} = 
\begin{bmatrix}
1/2\\
1/2
\end{bmatrix}\\
P_{Y_2Y_1}=
\begin{bmatrix}
1/2&0\\
1/2&1/2\\
0&1/2
\end{bmatrix}\\
P_{Y_3Y_2}=
\begin{bmatrix}
1/2&0&0\\
1/2&1/2&0\\
0&1/2&1/2\\
0&0&1/2
\end{bmatrix}\\
\vdots
\end{align}
where we slightly abuse notation to index rows and columns by outcomes of the random variable $Y_{k+1}$ and $Y_k$, respectively. The number of paths $\#_{Y_N}$ for each outcome of the random variable $Y_N$ can then be obtained simply as
\begin{align}
    \#_{Y_N} = P_{Y_NY_{N-1}} P_{Y_{N-1}Y_{N-2}} \dots P_{Y_0}
\end{align}


\begin{thebibliography}{75}%
\makeatletter
\providecommand \@ifxundefined [1]{%
 \@ifx{#1\undefined}
}%
\providecommand \@ifnum [1]{%
 \ifnum #1\expandafter \@firstoftwo
 \else \expandafter \@secondoftwo
 \fi
}%
\providecommand \@ifx [1]{%
 \ifx #1\expandafter \@firstoftwo
 \else \expandafter \@secondoftwo
 \fi
}%
\providecommand \natexlab [1]{#1}%
\providecommand \enquote  [1]{``#1''}%
\providecommand \bibnamefont  [1]{#1}%
\providecommand \bibfnamefont [1]{#1}%
\providecommand \citenamefont [1]{#1}%
\providecommand \href@noop [0]{\@secondoftwo}%
\providecommand \href [0]{\begingroup \@sanitize@url \@href}%
\providecommand \@href[1]{\@@startlink{#1}\@@href}%
\providecommand \@@href[1]{\endgroup#1\@@endlink}%
\providecommand \@sanitize@url [0]{\catcode `\\12\catcode `\$12\catcode
  `\&12\catcode `\#12\catcode `\^12\catcode `\_12\catcode `\%12\relax}%
\providecommand \@@startlink[1]{}%
\providecommand \@@endlink[0]{}%
\providecommand \url  [0]{\begingroup\@sanitize@url \@url }%
\providecommand \@url [1]{\endgroup\@href {#1}{\urlprefix }}%
\providecommand \urlprefix  [0]{URL }%
\providecommand \Eprint [0]{\href }%
\providecommand \doibase [0]{https://doi.org/}%
\providecommand \selectlanguage [0]{\@gobble}%
\providecommand \bibinfo  [0]{\@secondoftwo}%
\providecommand \bibfield  [0]{\@secondoftwo}%
\providecommand \translation [1]{[#1]}%
\providecommand \BibitemOpen [0]{}%
\providecommand \bibitemStop [0]{}%
\providecommand \bibitemNoStop [0]{.\EOS\space}%
\providecommand \EOS [0]{\spacefactor3000\relax}%
\providecommand \BibitemShut  [1]{\csname bibitem#1\endcsname}%
\let\auto@bib@innerbib\@empty
%</preamble>
\bibitem [{\citenamefont {Hull}(2021)}]{hull2021}%
  \BibitemOpen
  \bibfield  {author} {\bibinfo {author} {\bibfnamefont {J.}~\bibnamefont
  {Hull}},\ }\href {https://elibrary.pearson.de/book/99.150005/9781292410623}
  {\emph {\bibinfo {title} {{Options, Futures, and Other Derivatives Global
  Edition}}}}\ (\bibinfo  {publisher} {Pearson Deutschland},\ \bibinfo {year}
  {2021})\ p.\ \bibinfo {pages} {880}\BibitemShut {NoStop}%
\bibitem [{\citenamefont {Black}\ and\ \citenamefont
  {Scholes}(1973)}]{black1973}%
  \BibitemOpen
  \bibfield  {author} {\bibinfo {author} {\bibfnamefont {F.}~\bibnamefont
  {Black}}\ and\ \bibinfo {author} {\bibfnamefont {M.}~\bibnamefont
  {Scholes}},\ }\bibfield  {title} {\bibinfo {title} {The pricing of options
  and corporate liabilities},\ }\href@noop {} {\bibfield  {journal} {\bibinfo
  {journal} {Journal of political economy}\ }\textbf {\bibinfo {volume} {81}},\
  \bibinfo {pages} {637} (\bibinfo {year} {1973})}\BibitemShut {NoStop}%
\bibitem [{\citenamefont {Cox}\ \emph {et~al.}(1979)\citenamefont {Cox},
  \citenamefont {Ross},\ and\ \citenamefont {Rubinstein}}]{cox1979}%
  \BibitemOpen
  \bibfield  {author} {\bibinfo {author} {\bibfnamefont {J.~C.}\ \bibnamefont
  {Cox}}, \bibinfo {author} {\bibfnamefont {S.~A.}\ \bibnamefont {Ross}},\ and\
  \bibinfo {author} {\bibfnamefont {M.}~\bibnamefont {Rubinstein}},\ }\bibfield
   {title} {\bibinfo {title} {Option pricing: A simplified approach},\
  }\href@noop {} {\bibfield  {journal} {\bibinfo  {journal} {Journal of
  financial Economics}\ }\textbf {\bibinfo {volume} {7}},\ \bibinfo {pages}
  {229} (\bibinfo {year} {1979})}\BibitemShut {NoStop}%
\bibitem [{\citenamefont {Rendleman}(1979)}]{rendleman1979}%
  \BibitemOpen
  \bibfield  {author} {\bibinfo {author} {\bibfnamefont {R.~J.}\ \bibnamefont
  {Rendleman}},\ }\bibfield  {title} {\bibinfo {title} {Two-state option
  pricing},\ }\href@noop {} {\bibfield  {journal} {\bibinfo  {journal} {The
  Journal of Finance}\ }\textbf {\bibinfo {volume} {34}},\ \bibinfo {pages}
  {1093} (\bibinfo {year} {1979})}\BibitemShut {NoStop}%
\bibitem [{\citenamefont {Buchen}(2012)}]{Buchen2012}%
  \BibitemOpen
  \bibfield  {author} {\bibinfo {author} {\bibfnamefont {P.}~\bibnamefont
  {Buchen}},\ }\href {https://doi.org/10.1201/b11589} {\emph {\bibinfo {title}
  {An Introduction to Exotic Option Pricing}}}\ (\bibinfo  {publisher} {Chapman
  and Hall/CRC},\ \bibinfo {year} {2012})\BibitemShut {NoStop}%
\bibitem [{\citenamefont {Ruf}\ and\ \citenamefont {Wang}(2019)}]{Ruf2019}%
  \BibitemOpen
  \bibfield  {author} {\bibinfo {author} {\bibfnamefont {J.}~\bibnamefont
  {Ruf}}\ and\ \bibinfo {author} {\bibfnamefont {W.}~\bibnamefont {Wang}},\
  }\bibfield  {title} {\bibinfo {title} {Neural networks for option pricing and
  hedging: A literature review},\ }\bibfield  {journal} {\bibinfo  {journal}
  {SSRN Electronic Journal}\ }\href {https://doi.org/10.2139/ssrn.3486363}
  {10.2139/ssrn.3486363} (\bibinfo {year} {2019})\BibitemShut {NoStop}%
\bibitem [{\citenamefont {Glau}\ \emph {et~al.}(2020)\citenamefont {Glau},
  \citenamefont {Kressner},\ and\ \citenamefont {Statti}}]{Glau2020}%
  \BibitemOpen
  \bibfield  {author} {\bibinfo {author} {\bibfnamefont {K.}~\bibnamefont
  {Glau}}, \bibinfo {author} {\bibfnamefont {D.}~\bibnamefont {Kressner}},\
  and\ \bibinfo {author} {\bibfnamefont {F.}~\bibnamefont {Statti}},\
  }\bibfield  {title} {\bibinfo {title} {Low-rank tensor approximation for
  chebyshev interpolation in parametric option pricing},\ }\href
  {https://doi.org/10.1137/19M1244172} {\bibfield  {journal} {\bibinfo
  {journal} {SIAM Journal on Financial Mathematics}\ }\textbf {\bibinfo
  {volume} {11}},\ \bibinfo {pages} {897} (\bibinfo {year} {2020})},\ \Eprint
  {https://arxiv.org/abs/https://doi.org/10.1137/19M1244172}
  {https://doi.org/10.1137/19M1244172} \BibitemShut {NoStop}%
\bibitem [{\citenamefont {Sakurai}\ \emph {et~al.}(2024)\citenamefont
  {Sakurai}, \citenamefont {Takahashi},\ and\ \citenamefont
  {Miyamoto}}]{SAKURAI2024}%
  \BibitemOpen
  \bibfield  {author} {\bibinfo {author} {\bibfnamefont {R.}~\bibnamefont
  {Sakurai}}, \bibinfo {author} {\bibfnamefont {H.}~\bibnamefont {Takahashi}},\
  and\ \bibinfo {author} {\bibfnamefont {K.}~\bibnamefont {Miyamoto}},\
  }\bibfield  {title} {\bibinfo {title} {Learning fourier-based parametric
  option pricing with tensor trains},\ }\href
  {https://doi.org/10.11517/jsaisigtwo.2024.FIN-033_213} {\bibfield  {journal}
  {\bibinfo  {journal} {JSAI Technical Report, Type 2 SIG}\ }\textbf {\bibinfo
  {volume} {2024}},\ \bibinfo {pages} {213} (\bibinfo {year}
  {2024})}\BibitemShut {NoStop}%
\bibitem [{\citenamefont {Kobayashi}\ \emph {et~al.}(2024)\citenamefont
  {Kobayashi}, \citenamefont {Suimon},\ and\ \citenamefont
  {Miyamoto}}]{kobayashi2024}%
  \BibitemOpen
  \bibfield  {author} {\bibinfo {author} {\bibfnamefont {N.}~\bibnamefont
  {Kobayashi}}, \bibinfo {author} {\bibfnamefont {Y.}~\bibnamefont {Suimon}},\
  and\ \bibinfo {author} {\bibfnamefont {K.}~\bibnamefont {Miyamoto}},\ }\href
  {https://arxiv.org/abs/2402.17148} {\bibinfo {title} {Time series generation
  for option pricing on quantum computers using tensor network}} (\bibinfo
  {year} {2024}),\ \Eprint {https://arxiv.org/abs/2402.17148} {arXiv:2402.17148
  [quant-ph]} \BibitemShut {NoStop}%
\bibitem [{\citenamefont {Patel}\ \emph
  {et~al.}(2022{\natexlab{a}})\citenamefont {Patel}, \citenamefont {Hsing},
  \citenamefont {Sahin}, \citenamefont {Jahromi}, \citenamefont {Palmer},
  \citenamefont {Sharma}, \citenamefont {Michel}, \citenamefont {Porte},
  \citenamefont {Abid}, \citenamefont {Aubert}, \citenamefont {Castellani},
  \citenamefont {Lee}, \citenamefont {Mugel},\ and\ \citenamefont
  {Orus}}]{Patel2022}%
  \BibitemOpen
  \bibfield  {author} {\bibinfo {author} {\bibfnamefont {R.}~\bibnamefont
  {Patel}}, \bibinfo {author} {\bibfnamefont {C.-W.}\ \bibnamefont {Hsing}},
  \bibinfo {author} {\bibfnamefont {S.}~\bibnamefont {Sahin}}, \bibinfo
  {author} {\bibfnamefont {S.~S.}\ \bibnamefont {Jahromi}}, \bibinfo {author}
  {\bibfnamefont {S.}~\bibnamefont {Palmer}}, \bibinfo {author} {\bibfnamefont
  {S.}~\bibnamefont {Sharma}}, \bibinfo {author} {\bibfnamefont
  {C.}~\bibnamefont {Michel}}, \bibinfo {author} {\bibfnamefont
  {V.}~\bibnamefont {Porte}}, \bibinfo {author} {\bibfnamefont
  {M.}~\bibnamefont {Abid}}, \bibinfo {author} {\bibfnamefont {S.}~\bibnamefont
  {Aubert}}, \bibinfo {author} {\bibfnamefont {P.}~\bibnamefont {Castellani}},
  \bibinfo {author} {\bibfnamefont {C.-G.}\ \bibnamefont {Lee}}, \bibinfo
  {author} {\bibfnamefont {S.}~\bibnamefont {Mugel}},\ and\ \bibinfo {author}
  {\bibfnamefont {R.}~\bibnamefont {Orus}},\ }\href
  {https://doi.org/10.48550/ARXIV.2208.02235} {\bibinfo {title}
  {Quantum-inspired tensor neural networks for partial differential equations}}
  (\bibinfo {year} {2022}{\natexlab{a}})\BibitemShut {NoStop}%
\bibitem [{\citenamefont {Kastoryano}\ and\ \citenamefont
  {Pancotti}(2022)}]{Kastoryano2022}%
  \BibitemOpen
  \bibfield  {author} {\bibinfo {author} {\bibfnamefont {M.~J.}\ \bibnamefont
  {Kastoryano}}\ and\ \bibinfo {author} {\bibfnamefont {N.}~\bibnamefont
  {Pancotti}},\ }\bibfield  {title} {\bibinfo {title} {A highly efficient
  tensor network algorithm for multi-asset fourier options pricing}\ }(\bibinfo
  {year} {2022})\BibitemShut {NoStop}%
\bibitem [{\citenamefont {Cassel}(2022)}]{Cassel2022}%
  \BibitemOpen
  \bibfield  {author} {\bibinfo {author} {\bibfnamefont {S.}~\bibnamefont
  {Cassel}},\ }\href {https://EconPapers.repec.org/RePEc:arx:papers:2202.09780}
  {\emph {\bibinfo {title} {Fast high-dimensional integration using tensor
  networks}}},\ \bibinfo {type} {Papers}\ (\bibinfo  {institution}
  {arXiv.org},\ \bibinfo {year} {2022})\BibitemShut {NoStop}%
\bibitem [{\citenamefont {Antonov}\ and\ \citenamefont
  {Piterbarg}(2021)}]{antonov_alternatives_2021}%
  \BibitemOpen
  \bibfield  {author} {\bibinfo {author} {\bibfnamefont {A.}~\bibnamefont
  {Antonov}}\ and\ \bibinfo {author} {\bibfnamefont {V.}~\bibnamefont
  {Piterbarg}},\ }\href {https://doi.org/10.2139/ssrn.3958331}
  {{\selectlanguage {english}\bibinfo {title} {Alternatives to {Deep} {Neural}
  {Networks} in {Finance}}}} (\bibinfo {year} {2021})\BibitemShut {NoStop}%
\bibitem [{\citenamefont {Xu}\ \emph {et~al.}(2021)\citenamefont {Xu},
  \citenamefont {Calvi},\ and\ \citenamefont {Mandic}}]{xu_tensor-train_2021}%
  \BibitemOpen
  \bibfield  {author} {\bibinfo {author} {\bibfnamefont {Y.~L.}\ \bibnamefont
  {Xu}}, \bibinfo {author} {\bibfnamefont {G.~G.}\ \bibnamefont {Calvi}},\ and\
  \bibinfo {author} {\bibfnamefont {D.~P.}\ \bibnamefont {Mandic}},\ }\href
  {http://arxiv.org/abs/2105.04983} {\bibinfo {title} {Tensor-{Train}
  {Recurrent} {Neural} {Networks} for {Interpretable} {Multi}-{Way} {Financial}
  {Forecasting}}} (\bibinfo {year} {2021}),\ \bibinfo {note} {arXiv:2105.04983
  [cs]}\BibitemShut {NoStop}%
\bibitem [{\citenamefont {Mugel}\ \emph {et~al.}(2020)\citenamefont {Mugel},
  \citenamefont {Lizaso},\ and\ \citenamefont {Orus}}]{mugel_use_2020}%
  \BibitemOpen
  \bibfield  {author} {\bibinfo {author} {\bibfnamefont {S.}~\bibnamefont
  {Mugel}}, \bibinfo {author} {\bibfnamefont {E.}~\bibnamefont {Lizaso}},\ and\
  \bibinfo {author} {\bibfnamefont {R.}~\bibnamefont {Orus}},\ }\href
  {https://doi.org/10.48550/arXiv.2010.01312} {\bibinfo {title} {Use {Cases} of
  {Quantum} {Optimization} for {Finance}}} (\bibinfo {year} {2020}),\ \bibinfo
  {note} {arXiv:2010.01312 [q-fin]}\BibitemShut {NoStop}%
\bibitem [{\citenamefont {Mugel}\ \emph {et~al.}(2022)\citenamefont {Mugel},
  \citenamefont {Kuchkovsky}, \citenamefont {Sanchez}, \citenamefont
  {Fernandez-Lorenzo}, \citenamefont {Luis-Hita}, \citenamefont {Lizaso},\ and\
  \citenamefont {Orus}}]{mugel_dynamic_2022}%
  \BibitemOpen
  \bibfield  {author} {\bibinfo {author} {\bibfnamefont {S.}~\bibnamefont
  {Mugel}}, \bibinfo {author} {\bibfnamefont {C.}~\bibnamefont {Kuchkovsky}},
  \bibinfo {author} {\bibfnamefont {E.}~\bibnamefont {Sanchez}}, \bibinfo
  {author} {\bibfnamefont {S.}~\bibnamefont {Fernandez-Lorenzo}}, \bibinfo
  {author} {\bibfnamefont {J.}~\bibnamefont {Luis-Hita}}, \bibinfo {author}
  {\bibfnamefont {E.}~\bibnamefont {Lizaso}},\ and\ \bibinfo {author}
  {\bibfnamefont {R.}~\bibnamefont {Orus}},\ }\bibfield  {title} {\bibinfo
  {title} {Dynamic {Portfolio} {Optimization} with {Real} {Datasets} {Using}
  {Quantum} {Processors} and {Quantum}-{Inspired} {Tensor} {Networks}},\ }\href
  {https://doi.org/10.1103/PhysRevResearch.4.013006} {\bibfield  {journal}
  {\bibinfo  {journal} {Physical Review Research}\ }\textbf {\bibinfo {volume}
  {4}},\ \bibinfo {pages} {013006} (\bibinfo {year} {2022})},\ \bibinfo {note}
  {arXiv:2007.00017 [quant-ph, q-fin]}\BibitemShut {NoStop}%
\bibitem [{\citenamefont {Patel}\ \emph
  {et~al.}(2022{\natexlab{b}})\citenamefont {Patel}, \citenamefont {Hsing},
  \citenamefont {Sahin}, \citenamefont {Palmer}, \citenamefont {Jahromi},
  \citenamefont {Sharma}, \citenamefont {Dominguez}, \citenamefont {Tziritas},
  \citenamefont {Michel}, \citenamefont {Porte}, \citenamefont {Abid},
  \citenamefont {Aubert}, \citenamefont {Castellani}, \citenamefont {Mugel},\
  and\ \citenamefont {Orus}}]{patel_quantum-inspired_2022}%
  \BibitemOpen
  \bibfield  {author} {\bibinfo {author} {\bibfnamefont {R.~G.}\ \bibnamefont
  {Patel}}, \bibinfo {author} {\bibfnamefont {C.-W.}\ \bibnamefont {Hsing}},
  \bibinfo {author} {\bibfnamefont {S.}~\bibnamefont {Sahin}}, \bibinfo
  {author} {\bibfnamefont {S.}~\bibnamefont {Palmer}}, \bibinfo {author}
  {\bibfnamefont {S.~S.}\ \bibnamefont {Jahromi}}, \bibinfo {author}
  {\bibfnamefont {S.}~\bibnamefont {Sharma}}, \bibinfo {author} {\bibfnamefont
  {T.}~\bibnamefont {Dominguez}}, \bibinfo {author} {\bibfnamefont
  {K.}~\bibnamefont {Tziritas}}, \bibinfo {author} {\bibfnamefont
  {C.}~\bibnamefont {Michel}}, \bibinfo {author} {\bibfnamefont
  {V.}~\bibnamefont {Porte}}, \bibinfo {author} {\bibfnamefont
  {M.}~\bibnamefont {Abid}}, \bibinfo {author} {\bibfnamefont {S.}~\bibnamefont
  {Aubert}}, \bibinfo {author} {\bibfnamefont {P.}~\bibnamefont {Castellani}},
  \bibinfo {author} {\bibfnamefont {S.}~\bibnamefont {Mugel}},\ and\ \bibinfo
  {author} {\bibfnamefont {R.}~\bibnamefont {Orus}},\ }\href
  {https://doi.org/10.48550/arXiv.2212.14076} {\bibinfo {title}
  {Quantum-{Inspired} {Tensor} {Neural} {Networks} for {Option} {Pricing}}}
  (\bibinfo {year} {2022}{\natexlab{b}}),\ \bibinfo {note} {arXiv:2212.14076
  [quant-ph, q-fin]}\BibitemShut {NoStop}%
\bibitem [{\citenamefont {Patel}\ \emph {et~al.}(2024)\citenamefont {Patel},
  \citenamefont {Dominguez}, \citenamefont {Dib}, \citenamefont {Palmer},
  \citenamefont {Cadarso}, \citenamefont {Contreras}, \citenamefont {Ratnani},
  \citenamefont {Casanova}, \citenamefont {Hernández-Santana}, \citenamefont
  {Álvaro Díaz-Fernández}, \citenamefont {Andrés}, \citenamefont
  {Luis-Hita}, \citenamefont {Sánchez-Martínez}, \citenamefont {Mugel},\ and\
  \citenamefont {Orus}}]{patel_application_2024}%
  \BibitemOpen
  \bibfield  {author} {\bibinfo {author} {\bibfnamefont {R.~G.}\ \bibnamefont
  {Patel}}, \bibinfo {author} {\bibfnamefont {T.}~\bibnamefont {Dominguez}},
  \bibinfo {author} {\bibfnamefont {M.}~\bibnamefont {Dib}}, \bibinfo {author}
  {\bibfnamefont {S.}~\bibnamefont {Palmer}}, \bibinfo {author} {\bibfnamefont
  {A.}~\bibnamefont {Cadarso}}, \bibinfo {author} {\bibfnamefont {F.~D.~L.}\
  \bibnamefont {Contreras}}, \bibinfo {author} {\bibfnamefont {A.}~\bibnamefont
  {Ratnani}}, \bibinfo {author} {\bibfnamefont {F.~G.}\ \bibnamefont
  {Casanova}}, \bibinfo {author} {\bibfnamefont {S.}~\bibnamefont
  {Hernández-Santana}}, \bibinfo {author} {\bibnamefont {Álvaro
  Díaz-Fernández}}, \bibinfo {author} {\bibfnamefont {E.}~\bibnamefont
  {Andrés}}, \bibinfo {author} {\bibfnamefont {J.}~\bibnamefont {Luis-Hita}},
  \bibinfo {author} {\bibfnamefont {E.}~\bibnamefont {Sánchez-Martínez}},
  \bibinfo {author} {\bibfnamefont {S.}~\bibnamefont {Mugel}},\ and\ \bibinfo
  {author} {\bibfnamefont {R.}~\bibnamefont {Orus}},\ }\href
  {https://arxiv.org/abs/2304.09750} {\bibinfo {title} {Application of tensor
  neural networks to pricing bermudan swaptions}} (\bibinfo {year} {2024}),\
  \Eprint {https://arxiv.org/abs/2304.09750} {arXiv:2304.09750 [q-fin.CP]}
  \BibitemShut {NoStop}%
\bibitem [{\citenamefont {Perez-Garcia}\ \emph {et~al.}(2006)\citenamefont
  {Perez-Garcia}, \citenamefont {Verstraete}, \citenamefont {Wolf},\ and\
  \citenamefont {Cirac}}]{perez-garcia_matrix_2006}%
  \BibitemOpen
  \bibfield  {author} {\bibinfo {author} {\bibfnamefont {D.}~\bibnamefont
  {Perez-Garcia}}, \bibinfo {author} {\bibfnamefont {F.}~\bibnamefont
  {Verstraete}}, \bibinfo {author} {\bibfnamefont {M.~M.}\ \bibnamefont
  {Wolf}},\ and\ \bibinfo {author} {\bibfnamefont {J.~I.}\ \bibnamefont
  {Cirac}},\ }\bibfield  {title} {\bibinfo {title} {Matrix {Product} {State}
  {Representations}},\ }\href {http://arxiv.org/abs/quant-ph/0608197}
  {\bibfield  {journal} {\bibinfo  {journal} {quant-ph/0608197}\ } (\bibinfo
  {year} {2006})},\ \bibinfo {note} {quantum Inf. Comput. 7, 401
  (2007)}\BibitemShut {NoStop}%
\bibitem [{\citenamefont {Verstraete}\ \emph {et~al.}(2008)\citenamefont
  {Verstraete}, \citenamefont {Murg},\ and\ \citenamefont
  {Cirac}}]{verstraete_matrix_2008}%
  \BibitemOpen
  \bibfield  {author} {\bibinfo {author} {\bibfnamefont {F.}~\bibnamefont
  {Verstraete}}, \bibinfo {author} {\bibfnamefont {V.}~\bibnamefont {Murg}},\
  and\ \bibinfo {author} {\bibfnamefont {J.}~\bibnamefont {Cirac}},\ }\bibfield
   {title} {\bibinfo {title} {Matrix product states, projected entangled pair
  states, and variational renormalization group methods for quantum spin
  systems},\ }\href {https://doi.org/10.1080/14789940801912366} {\bibfield
  {journal} {\bibinfo  {journal} {Advances in Physics}\ }\textbf {\bibinfo
  {volume} {57}},\ \bibinfo {pages} {143} (\bibinfo {year} {2008})},\ \bibinfo
  {note} {publisher: Taylor \& Francis \_eprint:
  https://doi.org/10.1080/14789940801912366}\BibitemShut {NoStop}%
\bibitem [{\citenamefont
  {Schollwoeck}(2011)}]{schollwoeck_density-matrix_2011}%
  \BibitemOpen
  \bibfield  {author} {\bibinfo {author} {\bibfnamefont {U.}~\bibnamefont
  {Schollwoeck}},\ }\bibfield  {title} {\bibinfo {title} {The density-matrix
  renormalization group in the age of matrix product states},\ }\href
  {https://doi.org/10.1016/j.aop.2010.09.012} {\bibfield  {journal} {\bibinfo
  {journal} {Annals of Physics}\ }\textbf {\bibinfo {volume} {326}},\ \bibinfo
  {pages} {96} (\bibinfo {year} {2011})},\ \bibinfo {note} {arXiv:
  1008.3477}\BibitemShut {NoStop}%
\bibitem [{\citenamefont {Evenbly}\ and\ \citenamefont
  {Vidal}(2009)}]{evenbly_algorithms_2009}%
  \BibitemOpen
  \bibfield  {author} {\bibinfo {author} {\bibfnamefont {G.}~\bibnamefont
  {Evenbly}}\ and\ \bibinfo {author} {\bibfnamefont {G.}~\bibnamefont
  {Vidal}},\ }\bibfield  {title} {\bibinfo {title} {Algorithms for entanglement
  renormalization},\ }\href {https://doi.org/10.1103/PhysRevB.79.144108}
  {\bibfield  {journal} {\bibinfo  {journal} {Physical Review B}\ }\textbf
  {\bibinfo {volume} {79}},\ \bibinfo {pages} {144108} (\bibinfo {year}
  {2009})}\BibitemShut {NoStop}%
\bibitem [{\citenamefont {Chan}\ \emph {et~al.}(2016)\citenamefont {Chan},
  \citenamefont {Keselman}, \citenamefont {Nakatani}, \citenamefont {Li},\ and\
  \citenamefont {White}}]{chan_matrix_2016}%
  \BibitemOpen
  \bibfield  {author} {\bibinfo {author} {\bibfnamefont {G.~K.-L.}\
  \bibnamefont {Chan}}, \bibinfo {author} {\bibfnamefont {A.}~\bibnamefont
  {Keselman}}, \bibinfo {author} {\bibfnamefont {N.}~\bibnamefont {Nakatani}},
  \bibinfo {author} {\bibfnamefont {Z.}~\bibnamefont {Li}},\ and\ \bibinfo
  {author} {\bibfnamefont {S.~R.}\ \bibnamefont {White}},\ }\bibfield  {title}
  {\bibinfo {title} {Matrix product operators, matrix product states, and ab
  initio density matrix renormalization group algorithms},\ }\href
  {https://doi.org/10.1063/1.4955108} {\bibfield  {journal} {\bibinfo
  {journal} {The Journal of Chemical Physics}\ }\textbf {\bibinfo {volume}
  {145}},\ \bibinfo {pages} {014102} (\bibinfo {year} {2016})}\BibitemShut
  {NoStop}%
\bibitem [{\citenamefont {Evenbly}(2022)}]{evenbly_practical_2022}%
  \BibitemOpen
  \bibfield  {author} {\bibinfo {author} {\bibfnamefont {G.}~\bibnamefont
  {Evenbly}},\ }\href {https://doi.org/10.48550/arXiv.2202.02138} {\bibinfo
  {title} {A {Practical} {Guide} to the {Numerical} {Implementation} of
  {Tensor} {Networks} {I}: {Contractions}, {Decompositions} and {Gauge}
  {Freedom}}} (\bibinfo {year} {2022}),\ \bibinfo {note}
  {arXiv:2202.02138}\BibitemShut {NoStop}%
\bibitem [{\citenamefont {Orus}(2013)}]{orus_practical_2013}%
  \BibitemOpen
  \bibfield  {author} {\bibinfo {author} {\bibfnamefont {R.}~\bibnamefont
  {Orus}},\ }\bibfield  {title} {\bibinfo {title} {A {Practical} {Introduction}
  to {Tensor} {Networks}: {Matrix} {Product} {States} and {Projected}
  {Entangled} {Pair} {States}},\ }\href {http://arxiv.org/abs/1306.2164}
  {\bibfield  {journal} {\bibinfo  {journal} {arXiv:1306.2164 [cond-mat,
  physics:hep-lat, physics:hep-th, physics:quant-ph]}\ } (\bibinfo {year}
  {2013})}\BibitemShut {NoStop}%
\bibitem [{\citenamefont {Cirac}\ \emph {et~al.}(2021)\citenamefont {Cirac},
  \citenamefont {Pérez-García}, \citenamefont {Schuch},\ and\ \citenamefont
  {Verstraete}}]{cirac_matrix_2021}%
  \BibitemOpen
  \bibfield  {author} {\bibinfo {author} {\bibfnamefont {J.~I.}\ \bibnamefont
  {Cirac}}, \bibinfo {author} {\bibfnamefont {D.}~\bibnamefont
  {Pérez-García}}, \bibinfo {author} {\bibfnamefont {N.}~\bibnamefont
  {Schuch}},\ and\ \bibinfo {author} {\bibfnamefont {F.}~\bibnamefont
  {Verstraete}},\ }\bibfield  {title} {\bibinfo {title} {Matrix product states
  and projected entangled pair states: {Concepts}, symmetries, theorems},\
  }\href {https://doi.org/10.1103/RevModPhys.93.045003} {\bibfield  {journal}
  {\bibinfo  {journal} {Reviews of Modern Physics}\ }\textbf {\bibinfo {volume}
  {93}},\ \bibinfo {pages} {045003} (\bibinfo {year} {2021})},\ \bibinfo {note}
  {publisher: American Physical Society}\BibitemShut {NoStop}%
\bibitem [{\citenamefont {Wilson}(1975)}]{wilson_renormalization_1975}%
  \BibitemOpen
  \bibfield  {author} {\bibinfo {author} {\bibfnamefont {K.~G.}\ \bibnamefont
  {Wilson}},\ }\bibfield  {title} {\bibinfo {title} {The renormalization group:
  {Critical} phenomena and the {Kondo} problem},\ }\href
  {https://doi.org/10.1103/RevModPhys.47.773} {\bibfield  {journal} {\bibinfo
  {journal} {Reviews of Modern Physics}\ }\textbf {\bibinfo {volume} {47}},\
  \bibinfo {pages} {773} (\bibinfo {year} {1975})}\BibitemShut {NoStop}%
\bibitem [{\citenamefont {Affleck}\ \emph {et~al.}(1987)\citenamefont
  {Affleck}, \citenamefont {Kennedy}, \citenamefont {Lieb},\ and\ \citenamefont
  {Tasaki}}]{affleck_rigorous_1987}%
  \BibitemOpen
  \bibfield  {author} {\bibinfo {author} {\bibfnamefont {I.}~\bibnamefont
  {Affleck}}, \bibinfo {author} {\bibfnamefont {T.}~\bibnamefont {Kennedy}},
  \bibinfo {author} {\bibfnamefont {E.~H.}\ \bibnamefont {Lieb}},\ and\
  \bibinfo {author} {\bibfnamefont {H.}~\bibnamefont {Tasaki}},\ }\bibfield
  {title} {\bibinfo {title} {Rigorous results on valence-bond ground states in
  antiferromagnets},\ }\href {https://doi.org/10.1103/PhysRevLett.59.799}
  {\bibfield  {journal} {\bibinfo  {journal} {Physical Review Letters}\
  }\textbf {\bibinfo {volume} {59}},\ \bibinfo {pages} {799} (\bibinfo {year}
  {1987})}\BibitemShut {NoStop}%
\bibitem [{\citenamefont {Rommer}\ and\ \citenamefont
  {Östlund}(1997)}]{rommer_class_1997}%
  \BibitemOpen
  \bibfield  {author} {\bibinfo {author} {\bibfnamefont {S.}~\bibnamefont
  {Rommer}}\ and\ \bibinfo {author} {\bibfnamefont {S.}~\bibnamefont
  {Östlund}},\ }\bibfield  {title} {\bibinfo {title} {Class of ansatz wave
  functions for one-dimensional spin systems and their relation to the density
  matrix renormalization group},\ }\href
  {https://doi.org/10.1103/PhysRevB.55.2164} {\bibfield  {journal} {\bibinfo
  {journal} {Physical Review B}\ }\textbf {\bibinfo {volume} {55}},\ \bibinfo
  {pages} {2164} (\bibinfo {year} {1997})},\ \bibinfo {note} {copyright (C)
  2009 The American Physical Society; Please report any problems to
  prola@aps.org}\BibitemShut {NoStop}%
\bibitem [{\citenamefont {White}(1992)}]{white_density_1992}%
  \BibitemOpen
  \bibfield  {author} {\bibinfo {author} {\bibfnamefont {S.~R.}\ \bibnamefont
  {White}},\ }\bibfield  {title} {\bibinfo {title} {Density matrix formulation
  for quantum renormalization groups},\ }\href
  {https://doi.org/10.1103/PhysRevLett.69.2863} {\bibfield  {journal} {\bibinfo
   {journal} {Physical Review Letters}\ }\textbf {\bibinfo {volume} {69}},\
  \bibinfo {pages} {2863} (\bibinfo {year} {1992})}\BibitemShut {NoStop}%
\bibitem [{\citenamefont {White}(1993)}]{white_density-matrix_1993}%
  \BibitemOpen
  \bibfield  {author} {\bibinfo {author} {\bibfnamefont {S.~R.}\ \bibnamefont
  {White}},\ }\bibfield  {title} {\bibinfo {title} {Density-matrix algorithms
  for quantum renormalization groups},\ }\href
  {https://doi.org/10.1103/PhysRevB.48.10345} {\bibfield  {journal} {\bibinfo
  {journal} {Physical Review B}\ }\textbf {\bibinfo {volume} {48}},\ \bibinfo
  {pages} {10345} (\bibinfo {year} {1993})},\ \bibinfo {note} {copyright (C)
  2009 The American Physical Society; Please report any problems to
  prola@aps.org}\BibitemShut {NoStop}%
\bibitem [{\citenamefont {Verstraete}\ and\ \citenamefont
  {Cirac}(2004)}]{verstraete_renormalization_2004}%
  \BibitemOpen
  \bibfield  {author} {\bibinfo {author} {\bibfnamefont {F.}~\bibnamefont
  {Verstraete}}\ and\ \bibinfo {author} {\bibfnamefont {J.~I.}\ \bibnamefont
  {Cirac}},\ }\href {https://doi.org/10.48550/arXiv.cond-mat/0407066} {\bibinfo
  {title} {Renormalization algorithms for {Quantum}-{Many} {Body} {Systems} in
  two and higher dimensions}} (\bibinfo {year} {2004}),\ \bibinfo {note}
  {arXiv:cond-mat/0407066}\BibitemShut {NoStop}%
\bibitem [{\citenamefont {Vidal}(2007)}]{vidal_entanglement_2007}%
  \BibitemOpen
  \bibfield  {author} {\bibinfo {author} {\bibfnamefont {G.}~\bibnamefont
  {Vidal}},\ }\bibfield  {title} {\bibinfo {title} {Entanglement
  {Renormalization}},\ }\bibfield  {journal} {\bibinfo  {journal} {Physical
  Review Letters}\ }\textbf {\bibinfo {volume} {99}},\ \href
  {https://doi.org/10.1103/PhysRevLett.99.220405}
  {10.1103/PhysRevLett.99.220405} (\bibinfo {year} {2007})\BibitemShut
  {NoStop}%
\bibitem [{\citenamefont {Vidal}(2008)}]{vidal_class_2008}%
  \BibitemOpen
  \bibfield  {author} {\bibinfo {author} {\bibfnamefont {G.}~\bibnamefont
  {Vidal}},\ }\bibfield  {title} {\bibinfo {title} {Class of {Quantum}
  {Many}-{Body} {States} {That} {Can} {Be} {Efficiently} {Simulated}},\ }\href
  {https://doi.org/10.1103/PhysRevLett.101.110501} {\bibfield  {journal}
  {\bibinfo  {journal} {Physical Review Letters}\ }\textbf {\bibinfo {volume}
  {101}},\ \bibinfo {pages} {110501} (\bibinfo {year} {2008})}\BibitemShut
  {NoStop}%
\bibitem [{\citenamefont {Ö.~LEGEZA}\ and\ \citenamefont
  {HESS}(2003)}]{legeza_qc-dmrg_2003}%
  \BibitemOpen
  \bibfield  {author} {\bibinfo {author} {\bibfnamefont {J.~R.}\ \bibnamefont
  {Ö.~LEGEZA}}\ and\ \bibinfo {author} {\bibfnamefont {B.~A.}\ \bibnamefont
  {HESS}},\ }\bibfield  {title} {\bibinfo {title} {Qc-dmrg study of the
  ionic-neutral curve crossing of lif},\ }\href
  {https://doi.org/10.1080/0026897031000155625} {\bibfield  {journal} {\bibinfo
   {journal} {Molecular Physics}\ }\textbf {\bibinfo {volume} {101}},\ \bibinfo
  {pages} {2019} (\bibinfo {year} {2003})}\BibitemShut {NoStop}%
\bibitem [{\citenamefont {White}\ and\ \citenamefont
  {Martin}(1999)}]{white_ab_1999}%
  \BibitemOpen
  \bibfield  {author} {\bibinfo {author} {\bibfnamefont {S.~R.}\ \bibnamefont
  {White}}\ and\ \bibinfo {author} {\bibfnamefont {R.~L.}\ \bibnamefont
  {Martin}},\ }\bibfield  {title} {\bibinfo {title} {Ab initio quantum
  chemistry using the density matrix renormalization group},\ }\href
  {https://doi.org/10.1063/1.478295} {\bibfield  {journal} {\bibinfo  {journal}
  {The Journal of Chemical Physics}\ }\textbf {\bibinfo {volume} {110}},\
  \bibinfo {pages} {4127} (\bibinfo {year} {1999})},\ \bibinfo {note}
  {publisher: American Institute of Physics}\BibitemShut {NoStop}%
\bibitem [{\citenamefont {Chan}\ and\ \citenamefont
  {Sharma}(2011)}]{chan_density_2011}%
  \BibitemOpen
  \bibfield  {author} {\bibinfo {author} {\bibfnamefont {G.~K.-L.}\
  \bibnamefont {Chan}}\ and\ \bibinfo {author} {\bibfnamefont {S.}~\bibnamefont
  {Sharma}},\ }\bibfield  {title} {\bibinfo {title} {The {Density} {Matrix}
  {Renormalization} {Group} in {Quantum} {Chemistry}},\ }\href
  {https://doi.org/10.1146/annurev-physchem-032210-103338} {\bibfield
  {journal} {\bibinfo  {journal} {Annual Review of Physical Chemistry}\
  }\textbf {\bibinfo {volume} {62}},\ \bibinfo {pages} {465} (\bibinfo {year}
  {2011})}\BibitemShut {NoStop}%
\bibitem [{\citenamefont {García}\ \emph {et~al.}(2004)\citenamefont
  {García}, \citenamefont {Hallberg},\ and\ \citenamefont
  {Rozenberg}}]{garcia_dynamical_2004}%
  \BibitemOpen
  \bibfield  {author} {\bibinfo {author} {\bibfnamefont {D.~J.}\ \bibnamefont
  {García}}, \bibinfo {author} {\bibfnamefont {K.}~\bibnamefont {Hallberg}},\
  and\ \bibinfo {author} {\bibfnamefont {M.~J.}\ \bibnamefont {Rozenberg}},\
  }\bibfield  {title} {\bibinfo {title} {Dynamical {Mean} {Field} {Theory} with
  the {Density} {Matrix} {Renormalization} {Group}},\ }\href
  {https://doi.org/10.1103/PhysRevLett.93.246403} {\bibfield  {journal}
  {\bibinfo  {journal} {Physical Review Letters}\ }\textbf {\bibinfo {volume}
  {93}},\ \bibinfo {pages} {246403} (\bibinfo {year} {2004})},\ \bibinfo {note}
  {publisher: American Physical Society}\BibitemShut {NoStop}%
\bibitem [{\citenamefont {Ganahl}\ \emph {et~al.}(2014)\citenamefont {Ganahl},
  \citenamefont {Thunström}, \citenamefont {Verstraete}, \citenamefont
  {Held},\ and\ \citenamefont {Evertz}}]{ganahl_chebyshev_2014}%
  \BibitemOpen
  \bibfield  {author} {\bibinfo {author} {\bibfnamefont {M.}~\bibnamefont
  {Ganahl}}, \bibinfo {author} {\bibfnamefont {P.}~\bibnamefont {Thunström}},
  \bibinfo {author} {\bibfnamefont {F.}~\bibnamefont {Verstraete}}, \bibinfo
  {author} {\bibfnamefont {K.}~\bibnamefont {Held}},\ and\ \bibinfo {author}
  {\bibfnamefont {H.~G.}\ \bibnamefont {Evertz}},\ }\bibfield  {title}
  {\bibinfo {title} {Chebyshev expansion for impurity models using matrix
  product states},\ }\href {https://doi.org/10.1103/PhysRevB.90.045144}
  {\bibfield  {journal} {\bibinfo  {journal} {Physical Review B}\ }\textbf
  {\bibinfo {volume} {90}},\ \bibinfo {pages} {045144} (\bibinfo {year}
  {2014})},\ \bibinfo {note} {publisher: American Physical Society}\BibitemShut
  {NoStop}%
\bibitem [{\citenamefont {Ganahl}\ \emph {et~al.}(2015)\citenamefont {Ganahl},
  \citenamefont {Aichhorn}, \citenamefont {Evertz}, \citenamefont {Thunström},
  \citenamefont {Held},\ and\ \citenamefont
  {Verstraete}}]{ganahl_efficient_2015}%
  \BibitemOpen
  \bibfield  {author} {\bibinfo {author} {\bibfnamefont {M.}~\bibnamefont
  {Ganahl}}, \bibinfo {author} {\bibfnamefont {M.}~\bibnamefont {Aichhorn}},
  \bibinfo {author} {\bibfnamefont {H.~G.}\ \bibnamefont {Evertz}}, \bibinfo
  {author} {\bibfnamefont {P.}~\bibnamefont {Thunström}}, \bibinfo {author}
  {\bibfnamefont {K.}~\bibnamefont {Held}},\ and\ \bibinfo {author}
  {\bibfnamefont {F.}~\bibnamefont {Verstraete}},\ }\bibfield  {title}
  {\bibinfo {title} {Efficient {DMFT} impurity solver using real-time dynamics
  with matrix product states},\ }\href
  {https://doi.org/10.1103/PhysRevB.92.155132} {\bibfield  {journal} {\bibinfo
  {journal} {Physical Review B}\ }\textbf {\bibinfo {volume} {92}},\ \bibinfo
  {pages} {155132} (\bibinfo {year} {2015})},\ \bibinfo {note} {publisher:
  American Physical Society}\BibitemShut {NoStop}%
\bibitem [{\citenamefont {Bauernfeind}\ \emph {et~al.}(2017)\citenamefont
  {Bauernfeind}, \citenamefont {Zingl}, \citenamefont {Triebl}, \citenamefont
  {Aichhorn},\ and\ \citenamefont {Evertz}}]{bauernfeind_fork_2017}%
  \BibitemOpen
  \bibfield  {author} {\bibinfo {author} {\bibfnamefont {D.}~\bibnamefont
  {Bauernfeind}}, \bibinfo {author} {\bibfnamefont {M.}~\bibnamefont {Zingl}},
  \bibinfo {author} {\bibfnamefont {R.}~\bibnamefont {Triebl}}, \bibinfo
  {author} {\bibfnamefont {M.}~\bibnamefont {Aichhorn}},\ and\ \bibinfo
  {author} {\bibfnamefont {H.~G.}\ \bibnamefont {Evertz}},\ }\bibfield  {title}
  {\bibinfo {title} {Fork {Tensor}-{Product} {States}: {Efficient}
  {Multiorbital} {Real}-{Time} {DMFT} {Solver}},\ }\href
  {https://doi.org/10.1103/PhysRevX.7.031013} {\bibfield  {journal} {\bibinfo
  {journal} {Physical Review X}\ }\textbf {\bibinfo {volume} {7}},\ \bibinfo
  {pages} {031013} (\bibinfo {year} {2017})},\ \bibinfo {note} {publisher:
  American Physical Society}\BibitemShut {NoStop}%
\bibitem [{\citenamefont {Menczer}\ \emph {et~al.}(2024)\citenamefont
  {Menczer}, \citenamefont {Kapás}, \citenamefont {Werner},\ and\
  \citenamefont {Legeza}}]{menczer_two-dimensional_2024}%
  \BibitemOpen
  \bibfield  {author} {\bibinfo {author} {\bibfnamefont {A.}~\bibnamefont
  {Menczer}}, \bibinfo {author} {\bibfnamefont {K.}~\bibnamefont {Kapás}},
  \bibinfo {author} {\bibfnamefont {M.~A.}\ \bibnamefont {Werner}},\ and\
  \bibinfo {author} {\bibfnamefont {O.}~\bibnamefont {Legeza}},\ }\bibfield
  {title} {{\selectlanguage {english}\bibinfo {title} {Two-dimensional quantum
  lattice models via mode optimized hybrid {CPU}-{GPU} density matrix
  renormalization group method}},\ }\href
  {https://doi.org/10.1103/PhysRevB.109.195148} {\bibfield  {journal} {\bibinfo
   {journal} {Physical Review B}\ }\textbf {\bibinfo {volume} {109}},\ \bibinfo
  {pages} {195148} (\bibinfo {year} {2024})}\BibitemShut {NoStop}%
\bibitem [{\citenamefont {Wolf}\ \emph
  {et~al.}(2014{\natexlab{a}})\citenamefont {Wolf}, \citenamefont {McCulloch},
  \citenamefont {Parcollet},\ and\ \citenamefont
  {Schollwöck}}]{wolf_chebyshev_2014}%
  \BibitemOpen
  \bibfield  {author} {\bibinfo {author} {\bibfnamefont {F.~A.}\ \bibnamefont
  {Wolf}}, \bibinfo {author} {\bibfnamefont {I.~P.}\ \bibnamefont {McCulloch}},
  \bibinfo {author} {\bibfnamefont {O.}~\bibnamefont {Parcollet}},\ and\
  \bibinfo {author} {\bibfnamefont {U.}~\bibnamefont {Schollwöck}},\
  }\bibfield  {title} {\bibinfo {title} {Chebyshev matrix product state
  impurity solver for dynamical mean-field theory},\ }\href
  {https://doi.org/10.1103/PhysRevB.90.115124} {\bibfield  {journal} {\bibinfo
  {journal} {Physical Review B}\ }\textbf {\bibinfo {volume} {90}},\ \bibinfo
  {pages} {115124} (\bibinfo {year} {2014}{\natexlab{a}})},\ \bibinfo {note}
  {publisher: American Physical Society}\BibitemShut {NoStop}%
\bibitem [{\citenamefont {Wolf}\ \emph
  {et~al.}(2014{\natexlab{b}})\citenamefont {Wolf}, \citenamefont {McCulloch},\
  and\ \citenamefont {Schollwöck}}]{wolf_solving_2014}%
  \BibitemOpen
  \bibfield  {author} {\bibinfo {author} {\bibfnamefont {F.~A.}\ \bibnamefont
  {Wolf}}, \bibinfo {author} {\bibfnamefont {I.~P.}\ \bibnamefont
  {McCulloch}},\ and\ \bibinfo {author} {\bibfnamefont {U.}~\bibnamefont
  {Schollwöck}},\ }\bibfield  {title} {\bibinfo {title} {Solving
  nonequilibrium dynamical mean-field theory using matrix product states},\
  }\href {https://doi.org/10.1103/PhysRevB.90.235131} {\bibfield  {journal}
  {\bibinfo  {journal} {Physical Review B}\ }\textbf {\bibinfo {volume} {90}},\
  \bibinfo {pages} {235131} (\bibinfo {year} {2014}{\natexlab{b}})},\ \bibinfo
  {note} {publisher: American Physical Society}\BibitemShut {NoStop}%
\bibitem [{\citenamefont {Dolgov}\ \emph {et~al.}(2019)\citenamefont {Dolgov},
  \citenamefont {Anaya-Izquierdo}, \citenamefont {Fox},\ and\ \citenamefont
  {Scheichl}}]{dolgov_approximation_2019}%
  \BibitemOpen
  \bibfield  {author} {\bibinfo {author} {\bibfnamefont {S.}~\bibnamefont
  {Dolgov}}, \bibinfo {author} {\bibfnamefont {K.}~\bibnamefont
  {Anaya-Izquierdo}}, \bibinfo {author} {\bibfnamefont {C.}~\bibnamefont
  {Fox}},\ and\ \bibinfo {author} {\bibfnamefont {R.}~\bibnamefont
  {Scheichl}},\ }\href {http://arxiv.org/abs/1810.01212} {\bibinfo {title}
  {Approximation and sampling of multivariate probability distributions in the
  tensor train decomposition}} (\bibinfo {year} {2019}),\ \bibinfo {note}
  {arXiv:1810.01212 [cs, math, stat]}\BibitemShut {NoStop}%
\bibitem [{\citenamefont {Dolgov}\ and\ \citenamefont
  {Savostyanov}(2020)}]{DOLGOV2020}%
  \BibitemOpen
  \bibfield  {author} {\bibinfo {author} {\bibfnamefont {S.}~\bibnamefont
  {Dolgov}}\ and\ \bibinfo {author} {\bibfnamefont {D.}~\bibnamefont
  {Savostyanov}},\ }\bibfield  {title} {\bibinfo {title} {Parallel cross
  interpolation for high-precision calculation of high-dimensional integrals},\
  }\href {https://doi.org/https://doi.org/10.1016/j.cpc.2019.106869} {\bibfield
   {journal} {\bibinfo  {journal} {Computer Physics Communications}\ }\textbf
  {\bibinfo {volume} {246}},\ \bibinfo {pages} {106869} (\bibinfo {year}
  {2020})}\BibitemShut {NoStop}%
\bibitem [{\citenamefont {Oseledets}\ and\ \citenamefont
  {Tyrtyshnikov}(2010)}]{OSELEDETS2010}%
  \BibitemOpen
  \bibfield  {author} {\bibinfo {author} {\bibfnamefont {I.}~\bibnamefont
  {Oseledets}}\ and\ \bibinfo {author} {\bibfnamefont {E.}~\bibnamefont
  {Tyrtyshnikov}},\ }\bibfield  {title} {\bibinfo {title} {Tt-cross
  approximation for multidimensional arrays},\ }\href
  {https://doi.org/https://doi.org/10.1016/j.laa.2009.07.024} {\bibfield
  {journal} {\bibinfo  {journal} {Linear Algebra and its Applications}\
  }\textbf {\bibinfo {volume} {432}},\ \bibinfo {pages} {70} (\bibinfo {year}
  {2010})}\BibitemShut {NoStop}%
\bibitem [{\citenamefont {Glasser}\ \emph {et~al.}(2019)\citenamefont
  {Glasser}, \citenamefont {Sweke}, \citenamefont {Pancotti}, \citenamefont
  {Eisert},\ and\ \citenamefont {Cirac}}]{glasser_expressive_2019}%
  \BibitemOpen
  \bibfield  {author} {\bibinfo {author} {\bibfnamefont {I.}~\bibnamefont
  {Glasser}}, \bibinfo {author} {\bibfnamefont {R.}~\bibnamefont {Sweke}},
  \bibinfo {author} {\bibfnamefont {N.}~\bibnamefont {Pancotti}}, \bibinfo
  {author} {\bibfnamefont {J.}~\bibnamefont {Eisert}},\ and\ \bibinfo {author}
  {\bibfnamefont {J.~I.}\ \bibnamefont {Cirac}},\ }\bibfield  {title} {\bibinfo
  {title} {Expressive power of tensor-network factorizations for probabilistic
  modeling},\ }in\ \href@noop {} {\emph {\bibinfo {booktitle} {Proceedings of
  the 33rd {International} {Conference} on {Neural} {Information} {Processing}
  {Systems}}}},\ \bibinfo {series and number} {\bibinfo {number} {134}}\
  (\bibinfo  {publisher} {Curran Associates Inc.},\ \bibinfo {address} {Red
  Hook, NY, USA},\ \bibinfo {year} {2019})\ pp.\ \bibinfo {pages}
  {1498--1510}\BibitemShut {NoStop}%
\bibitem [{\citenamefont {Glasser}\ \emph {et~al.}(2020)\citenamefont
  {Glasser}, \citenamefont {Pancotti},\ and\ \citenamefont
  {Cirac}}]{glasser_probabilistic_2020}%
  \BibitemOpen
  \bibfield  {author} {\bibinfo {author} {\bibfnamefont {I.}~\bibnamefont
  {Glasser}}, \bibinfo {author} {\bibfnamefont {N.}~\bibnamefont {Pancotti}},\
  and\ \bibinfo {author} {\bibfnamefont {J.~I.}\ \bibnamefont {Cirac}},\
  }\bibfield  {title} {\bibinfo {title} {From {Probabilistic} {Graphical}
  {Models} to {Generalized} {Tensor} {Networks} for {Supervised} {Learning}},\
  }\href {https://doi.org/10.1109/ACCESS.2020.2986279} {\bibfield  {journal}
  {\bibinfo  {journal} {IEEE Access}\ }\textbf {\bibinfo {volume} {8}},\
  \bibinfo {pages} {68169} (\bibinfo {year} {2020})},\ \bibinfo {note}
  {conference Name: IEEE Access}\BibitemShut {NoStop}%
\bibitem [{\citenamefont {Cichocki}\ \emph {et~al.}(2016)\citenamefont
  {Cichocki}, \citenamefont {Lee}, \citenamefont {Oseledets}, \citenamefont
  {Phan}, \citenamefont {Zhao},\ and\ \citenamefont {Mandic}}]{Cichocki2016p1}%
  \BibitemOpen
  \bibfield  {author} {\bibinfo {author} {\bibfnamefont {A.}~\bibnamefont
  {Cichocki}}, \bibinfo {author} {\bibfnamefont {N.}~\bibnamefont {Lee}},
  \bibinfo {author} {\bibfnamefont {I.}~\bibnamefont {Oseledets}}, \bibinfo
  {author} {\bibfnamefont {A.-H.}\ \bibnamefont {Phan}}, \bibinfo {author}
  {\bibfnamefont {Q.}~\bibnamefont {Zhao}},\ and\ \bibinfo {author}
  {\bibfnamefont {D.~P.}\ \bibnamefont {Mandic}},\ }\bibfield  {title}
  {\bibinfo {title} {Tensor networks for dimensionality reduction and
  large-scale optimization: Part 1 low-rank tensor decompositions},\ }\href
  {https://doi.org/10.1561/2200000059} {\bibfield  {journal} {\bibinfo
  {journal} {Foundations and Trends{\textregistered} in Machine Learning}\
  }\textbf {\bibinfo {volume} {9}},\ \bibinfo {pages} {249–429} (\bibinfo
  {year} {2016})}\BibitemShut {NoStop}%
\bibitem [{\citenamefont {Cichocki}\ \emph {et~al.}(2017)\citenamefont
  {Cichocki}, \citenamefont {Lee}, \citenamefont {Oseledets}, \citenamefont
  {Phan}, \citenamefont {Zhao}, \citenamefont {Sugiyama},\ and\ \citenamefont
  {Mandic}}]{Cichocki2017p2}%
  \BibitemOpen
  \bibfield  {author} {\bibinfo {author} {\bibfnamefont {A.}~\bibnamefont
  {Cichocki}}, \bibinfo {author} {\bibfnamefont {N.}~\bibnamefont {Lee}},
  \bibinfo {author} {\bibfnamefont {I.}~\bibnamefont {Oseledets}}, \bibinfo
  {author} {\bibfnamefont {A.-H.}\ \bibnamefont {Phan}}, \bibinfo {author}
  {\bibfnamefont {Q.}~\bibnamefont {Zhao}}, \bibinfo {author} {\bibfnamefont
  {M.}~\bibnamefont {Sugiyama}},\ and\ \bibinfo {author} {\bibfnamefont
  {D.~P.}\ \bibnamefont {Mandic}},\ }\bibfield  {title} {\bibinfo {title}
  {Tensor networks for dimensionality reduction and large-scale optimization:
  Part 2 applications and future perspectives},\ }\href
  {https://doi.org/10.1561/2200000067} {\bibfield  {journal} {\bibinfo
  {journal} {Foundations and Trends{\textregistered} in Machine Learning}\
  }\textbf {\bibinfo {volume} {9}},\ \bibinfo {pages} {249–429} (\bibinfo
  {year} {2017})}\BibitemShut {NoStop}%
\bibitem [{\citenamefont {Goeßmann}\ \emph {et~al.}(2020)\citenamefont
  {Goeßmann}, \citenamefont {Götte}, \citenamefont {Roth}, \citenamefont
  {Sweke}, \citenamefont {Kutyniok},\ and\ \citenamefont
  {Eisert}}]{goesmann_tensor_2020}%
  \BibitemOpen
  \bibfield  {author} {\bibinfo {author} {\bibfnamefont {A.}~\bibnamefont
  {Goeßmann}}, \bibinfo {author} {\bibfnamefont {M.}~\bibnamefont {Götte}},
  \bibinfo {author} {\bibfnamefont {I.}~\bibnamefont {Roth}}, \bibinfo {author}
  {\bibfnamefont {R.}~\bibnamefont {Sweke}}, \bibinfo {author} {\bibfnamefont
  {G.}~\bibnamefont {Kutyniok}},\ and\ \bibinfo {author} {\bibfnamefont
  {J.}~\bibnamefont {Eisert}},\ }\href
  {https://doi.org/10.48550/arXiv.2002.12388} {\bibinfo {title} {Tensor network
  approaches for learning non-linear dynamical laws}} (\bibinfo {year}
  {2020}),\ \bibinfo {note} {arXiv:2002.12388 [quant-ph, stat]}\BibitemShut
  {NoStop}%
\bibitem [{\citenamefont {Ali}\ \emph {et~al.}(2024)\citenamefont {Ali},
  \citenamefont {Leceta},\ and\ \citenamefont {Rubio}}]{ali_anomaly_2024}%
  \BibitemOpen
  \bibfield  {author} {\bibinfo {author} {\bibfnamefont {A.~M.}\ \bibnamefont
  {Ali}}, \bibinfo {author} {\bibfnamefont {A.~M. F.~d.}\ \bibnamefont
  {Leceta}},\ and\ \bibinfo {author} {\bibfnamefont {J.~L.}\ \bibnamefont
  {Rubio}},\ }\href {http://arxiv.org/abs/2409.15030} {\bibinfo {title}
  {Anomaly {Detection} from a {Tensor} {Train} {Perspective}}} (\bibinfo {year}
  {2024}),\ \bibinfo {note} {arXiv:2409.15030}\BibitemShut {NoStop}%
\bibitem [{\citenamefont {Wang}\ \emph {et~al.}(2020)\citenamefont {Wang},
  \citenamefont {Roberts}, \citenamefont {Vidal},\ and\ \citenamefont
  {Leichenauer}}]{wang_anomaly_2020}%
  \BibitemOpen
  \bibfield  {author} {\bibinfo {author} {\bibfnamefont {J.}~\bibnamefont
  {Wang}}, \bibinfo {author} {\bibfnamefont {C.}~\bibnamefont {Roberts}},
  \bibinfo {author} {\bibfnamefont {G.}~\bibnamefont {Vidal}},\ and\ \bibinfo
  {author} {\bibfnamefont {S.}~\bibnamefont {Leichenauer}},\ }\href
  {https://doi.org/10.48550/arXiv.2006.02516} {\bibinfo {title} {Anomaly
  {Detection} with {Tensor} {Networks}}} (\bibinfo {year} {2020}),\ \bibinfo
  {note} {arXiv:2006.02516 [quant-ph, stat]}\BibitemShut {NoStop}%
\bibitem [{\citenamefont {Stoudenmire}(2018)}]{stoudenmire_learning_2018}%
  \BibitemOpen
  \bibfield  {author} {\bibinfo {author} {\bibfnamefont {E.~M.}\ \bibnamefont
  {Stoudenmire}},\ }\bibfield  {title} {\bibinfo {title} {Learning {Relevant}
  {Features} of {Data} with {Multi}-scale {Tensor} {Networks}},\ }\href
  {https://doi.org/10.1088/2058-9565/aaba1a} {\bibfield  {journal} {\bibinfo
  {journal} {Quantum Science and Technology}\ }\textbf {\bibinfo {volume}
  {3}},\ \bibinfo {pages} {034003} (\bibinfo {year} {2018})},\ \bibinfo {note}
  {arXiv:1801.00315 [cond-mat, stat]}\BibitemShut {NoStop}%
\bibitem [{\citenamefont {Wang}\ \emph {et~al.}(2018)\citenamefont {Wang},
  \citenamefont {Aggarwal},\ and\ \citenamefont {Aeron}}]{wang_principal_2018}%
  \BibitemOpen
  \bibfield  {author} {\bibinfo {author} {\bibfnamefont {W.}~\bibnamefont
  {Wang}}, \bibinfo {author} {\bibfnamefont {V.}~\bibnamefont {Aggarwal}},\
  and\ \bibinfo {author} {\bibfnamefont {S.}~\bibnamefont {Aeron}},\ }\href
  {http://arxiv.org/abs/1803.05026} {\bibinfo {title} {Principal {Component}
  {Analysis} with {Tensor} {Train} {Subspace}}} (\bibinfo {year} {2018}),\
  \bibinfo {note} {arXiv:1803.05026 [cs, math]}\BibitemShut {NoStop}%
\bibitem [{\citenamefont {Lu}\ \emph {et~al.}(2021)\citenamefont {Lu},
  \citenamefont {Kanász-Nagy}, \citenamefont {Kukuljan},\ and\ \citenamefont
  {Cirac}}]{lu_tensor_2021}%
  \BibitemOpen
  \bibfield  {author} {\bibinfo {author} {\bibfnamefont {S.}~\bibnamefont
  {Lu}}, \bibinfo {author} {\bibfnamefont {M.}~\bibnamefont {Kanász-Nagy}},
  \bibinfo {author} {\bibfnamefont {I.}~\bibnamefont {Kukuljan}},\ and\
  \bibinfo {author} {\bibfnamefont {J.~I.}\ \bibnamefont {Cirac}},\ }\href
  {https://doi.org/10.48550/arXiv.2103.06872} {\bibinfo {title} {Tensor
  networks and efficient descriptions of classical data}} (\bibinfo {year}
  {2021}),\ \bibinfo {note} {arXiv:2103.06872}\BibitemShut {NoStop}%
\bibitem [{\citenamefont {Liu}\ \emph {et~al.}(2023)\citenamefont {Liu},
  \citenamefont {Li}, \citenamefont {Zhang},\ and\ \citenamefont
  {Zhang}}]{liu_tensor_2023}%
  \BibitemOpen
  \bibfield  {author} {\bibinfo {author} {\bibfnamefont {J.}~\bibnamefont
  {Liu}}, \bibinfo {author} {\bibfnamefont {S.}~\bibnamefont {Li}}, \bibinfo
  {author} {\bibfnamefont {J.}~\bibnamefont {Zhang}},\ and\ \bibinfo {author}
  {\bibfnamefont {P.}~\bibnamefont {Zhang}},\ }\bibfield  {title} {\bibinfo
  {title} {Tensor networks for unsupervised machine learning},\ }\href
  {https://doi.org/10.1103/PhysRevE.107.L012103} {\bibfield  {journal}
  {\bibinfo  {journal} {Physical Review E}\ }\textbf {\bibinfo {volume}
  {107}},\ \bibinfo {pages} {L012103} (\bibinfo {year} {2023})},\ \bibinfo
  {note} {arXiv:2106.12974 [cond-mat, physics:quant-ph, stat]}\BibitemShut
  {NoStop}%
\bibitem [{\citenamefont {Konstantinidis}\ \emph {et~al.}(2021)\citenamefont
  {Konstantinidis}, \citenamefont {Xu}, \citenamefont {Mandic}, \citenamefont
  {Zhao},\ and\ \citenamefont {Team}}]{konstantinidis_bayesian_nodate}%
  \BibitemOpen
  \bibfield  {author} {\bibinfo {author} {\bibfnamefont {K.}~\bibnamefont
  {Konstantinidis}}, \bibinfo {author} {\bibfnamefont {Y.~L.}\ \bibnamefont
  {Xu}}, \bibinfo {author} {\bibfnamefont {D.~P.}\ \bibnamefont {Mandic}},
  \bibinfo {author} {\bibfnamefont {Q.}~\bibnamefont {Zhao}},\ and\ \bibinfo
  {author} {\bibfnamefont {T.~L.}\ \bibnamefont {Team}},\ }\bibfield  {title}
  {\bibinfo {title} {Bayesian tensor networks with structured posteriors},\
  }in\ \href@noop {} {\emph {\bibinfo {booktitle} {the Second Workshop on
  Quantum Tensor Networks in Machine Learning, 35th Conference on Neural
  Information Processing Systems (NIPS)}}}\ (\bibinfo {year} {2021})\
  p.~\bibinfo {pages} {22}\BibitemShut {NoStop}%
\bibitem [{\citenamefont {Kirstein}\ \emph {et~al.}(2022)\citenamefont
  {Kirstein}, \citenamefont {Sommer},\ and\ \citenamefont
  {Eigel}}]{kirstein_tensor-train_2022}%
  \BibitemOpen
  \bibfield  {author} {\bibinfo {author} {\bibfnamefont {M.}~\bibnamefont
  {Kirstein}}, \bibinfo {author} {\bibfnamefont {D.}~\bibnamefont {Sommer}},\
  and\ \bibinfo {author} {\bibfnamefont {M.}~\bibnamefont {Eigel}},\ }\bibfield
   {title} {\bibinfo {title} {Tensor-train kernel learning for gaussian
  processes},\ }in\ \href {https://proceedings.mlr.press/v179/kirstein22a.html}
  {\emph {\bibinfo {booktitle} {Proceedings of the Eleventh Symposium on
  Conformal and Probabilistic Prediction with Applications}}},\ \bibinfo
  {series} {Proceedings of Machine Learning Research}, Vol.\ \bibinfo {volume}
  {179},\ \bibinfo {editor} {edited by\ \bibinfo {editor} {\bibfnamefont
  {U.}~\bibnamefont {Johansson}}, \bibinfo {editor} {\bibfnamefont
  {H.}~\bibnamefont {Boström}}, \bibinfo {editor} {\bibfnamefont
  {K.}~\bibnamefont {An~Nguyen}}, \bibinfo {editor} {\bibfnamefont
  {Z.}~\bibnamefont {Luo}},\ and\ \bibinfo {editor} {\bibfnamefont
  {L.}~\bibnamefont {Carlsson}}}\ (\bibinfo  {publisher} {PMLR},\ \bibinfo
  {year} {2022})\ pp.\ \bibinfo {pages} {253--272}\BibitemShut {NoStop}%
\bibitem [{\citenamefont {Izmailov}\ \emph {et~al.}(2018)\citenamefont
  {Izmailov}, \citenamefont {Novikov},\ and\ \citenamefont
  {Kropotov}}]{izmailov_scalable_2018}%
  \BibitemOpen
  \bibfield  {author} {\bibinfo {author} {\bibfnamefont {P.}~\bibnamefont
  {Izmailov}}, \bibinfo {author} {\bibfnamefont {A.}~\bibnamefont {Novikov}},\
  and\ \bibinfo {author} {\bibfnamefont {D.}~\bibnamefont {Kropotov}},\ }\href
  {https://doi.org/10.48550/arXiv.1710.07324} {\bibinfo {title} {Scalable
  {Gaussian} {Processes} with {Billions} of {Inducing} {Inputs} via {Tensor}
  {Train} {Decomposition}}} (\bibinfo {year} {2018}),\ \bibinfo {note}
  {arXiv:1710.07324 [cs, stat]}\BibitemShut {NoStop}%
\bibitem [{\citenamefont {Han}\ \emph {et~al.}(2018)\citenamefont {Han},
  \citenamefont {Wang}, \citenamefont {Fan}, \citenamefont {Wang},\ and\
  \citenamefont {Zhang}}]{han_unsupervised_2018}%
  \BibitemOpen
  \bibfield  {author} {\bibinfo {author} {\bibfnamefont {Z.-Y.}\ \bibnamefont
  {Han}}, \bibinfo {author} {\bibfnamefont {J.}~\bibnamefont {Wang}}, \bibinfo
  {author} {\bibfnamefont {H.}~\bibnamefont {Fan}}, \bibinfo {author}
  {\bibfnamefont {L.}~\bibnamefont {Wang}},\ and\ \bibinfo {author}
  {\bibfnamefont {P.}~\bibnamefont {Zhang}},\ }\bibfield  {title} {\bibinfo
  {title} {Unsupervised {Generative} {Modeling} {Using} {Matrix} {Product}
  {States}},\ }\href {https://doi.org/10.1103/PhysRevX.8.031012} {\bibfield
  {journal} {\bibinfo  {journal} {Physical Review X}\ }\textbf {\bibinfo
  {volume} {8}},\ \bibinfo {pages} {031012} (\bibinfo {year} {2018})},\
  \bibinfo {note} {publisher: American Physical Society}\BibitemShut {NoStop}%
\bibitem [{\citenamefont {Peng}\ \emph {et~al.}(2023)\citenamefont {Peng},
  \citenamefont {Chen}, \citenamefont {Stoudenmire},\ and\ \citenamefont
  {Khoo}}]{peng_generative_2023}%
  \BibitemOpen
  \bibfield  {author} {\bibinfo {author} {\bibfnamefont {Y.}~\bibnamefont
  {Peng}}, \bibinfo {author} {\bibfnamefont {Y.}~\bibnamefont {Chen}}, \bibinfo
  {author} {\bibfnamefont {E.~M.}\ \bibnamefont {Stoudenmire}},\ and\ \bibinfo
  {author} {\bibfnamefont {Y.}~\bibnamefont {Khoo}},\ }\href
  {https://doi.org/10.48550/arXiv.2304.05305} {\bibinfo {title} {Generative
  {Modeling} via {Hierarchical} {Tensor} {Sketching}}} (\bibinfo {year}
  {2023}),\ \bibinfo {note} {arXiv:2304.05305}\BibitemShut {NoStop}%
\bibitem [{\citenamefont {Strashko}\ and\ \citenamefont
  {Stoudenmire}(2022)}]{strashko_generalization_2022}%
  \BibitemOpen
  \bibfield  {author} {\bibinfo {author} {\bibfnamefont {A.}~\bibnamefont
  {Strashko}}\ and\ \bibinfo {author} {\bibfnamefont {E.~M.}\ \bibnamefont
  {Stoudenmire}},\ }\href {https://doi.org/10.48550/arXiv.2208.04372} {\bibinfo
  {title} {Generalization and {Overfitting} in {Matrix} {Product} {State}
  {Machine} {Learning} {Architectures}}} (\bibinfo {year} {2022}),\ \bibinfo
  {note} {arXiv:2208.04372}\BibitemShut {NoStop}%
\bibitem [{\citenamefont {Jasra}\ and\ \citenamefont
  {Del~Moral}(2011)}]{Jasra2011}%
  \BibitemOpen
  \bibfield  {author} {\bibinfo {author} {\bibfnamefont {A.}~\bibnamefont
  {Jasra}}\ and\ \bibinfo {author} {\bibfnamefont {P.}~\bibnamefont
  {Del~Moral}},\ }\bibfield  {title} {\bibinfo {title} {Sequential monte carlo
  methods for option pricing},\ }\href
  {https://doi.org/10.1080/07362994.2011.548993} {\bibfield  {journal}
  {\bibinfo  {journal} {Stochastic Analysis and Applications}\ }\textbf
  {\bibinfo {volume} {29}},\ \bibinfo {pages} {292–316} (\bibinfo {year}
  {2011})}\BibitemShut {NoStop}%
\bibitem [{\citenamefont {Duffy}(2006)}]{Duffy2006}%
  \BibitemOpen
  \bibfield  {author} {\bibinfo {author} {\bibfnamefont {D.~J.}\ \bibnamefont
  {Duffy}},\ }\href {https://doi.org/10.1002/9781118673447} {\emph {\bibinfo
  {title} {Finite Difference Methods in Financial Engineering: A Partial
  Differential Equation Approach}}}\ (\bibinfo  {publisher} {Wiley},\ \bibinfo
  {year} {2006})\BibitemShut {NoStop}%
\bibitem [{\citenamefont {Bouchard}\ and\ \citenamefont
  {Warin}(2012)}]{Bouchard2012}%
  \BibitemOpen
  \bibfield  {author} {\bibinfo {author} {\bibfnamefont {B.}~\bibnamefont
  {Bouchard}}\ and\ \bibinfo {author} {\bibfnamefont {X.}~\bibnamefont
  {Warin}},\ }\bibinfo {title} {Monte-carlo valuation of american options:
  Facts and new algorithms to improve existing methods},\ in\ \href
  {https://doi.org/10.1007/978-3-642-25746-9_7} {\emph {\bibinfo {booktitle}
  {Numerical Methods in Finance}}}\ (\bibinfo  {publisher} {Springer Berlin
  Heidelberg},\ \bibinfo {year} {2012})\ p.\ \bibinfo {pages}
  {215–255}\BibitemShut {NoStop}%
\bibitem [{\citenamefont {Bridgeman}\ and\ \citenamefont
  {Chubb}(2017)}]{bridgeman_hand-waving_2017}%
  \BibitemOpen
  \bibfield  {author} {\bibinfo {author} {\bibfnamefont {J.~C.}\ \bibnamefont
  {Bridgeman}}\ and\ \bibinfo {author} {\bibfnamefont {C.~T.}\ \bibnamefont
  {Chubb}},\ }\bibfield  {title} {{\selectlanguage {english}\bibinfo {title}
  {Hand-waving and interpretive dance: an introductory course on tensor
  networks}},\ }\href {https://doi.org/10.1088/1751-8121/aa6dc3} {\bibfield
  {journal} {\bibinfo  {journal} {Journal of Physics A: Mathematical and
  Theoretical}\ }\textbf {\bibinfo {volume} {50}},\ \bibinfo {pages} {223001}
  (\bibinfo {year} {2017})}\BibitemShut {NoStop}%
\bibitem [{\citenamefont {Steinlechner}(2016)}]{steinlechner_riemannian_2016}%
  \BibitemOpen
  \bibfield  {author} {\bibinfo {author} {\bibfnamefont {M.}~\bibnamefont
  {Steinlechner}},\ }\bibfield  {title} {{\selectlanguage {english}\bibinfo
  {title} {Riemannian {Optimization} for {High}-{Dimensional} {Tensor}
  {Completion}}},\ }\href {https://doi.org/10.1137/15M1010506} {\bibfield
  {journal} {\bibinfo  {journal} {SIAM Journal on Scientific Computing}\
  }\textbf {\bibinfo {volume} {38}},\ \bibinfo {pages} {S461} (\bibinfo {year}
  {2016})}\BibitemShut {NoStop}%
\bibitem [{\citenamefont {Roberts}\ \emph {et~al.}(2019)\citenamefont
  {Roberts}, \citenamefont {Milsted}, \citenamefont {Ganahl}, \citenamefont
  {Zalcman}, \citenamefont {Fontaine}, \citenamefont {Zou}, \citenamefont
  {Hidary}, \citenamefont {Vidal},\ and\ \citenamefont
  {Leichenauer}}]{Roberts2019}%
  \BibitemOpen
  \bibfield  {author} {\bibinfo {author} {\bibfnamefont {C.}~\bibnamefont
  {Roberts}}, \bibinfo {author} {\bibfnamefont {A.}~\bibnamefont {Milsted}},
  \bibinfo {author} {\bibfnamefont {M.}~\bibnamefont {Ganahl}}, \bibinfo
  {author} {\bibfnamefont {A.}~\bibnamefont {Zalcman}}, \bibinfo {author}
  {\bibfnamefont {B.}~\bibnamefont {Fontaine}}, \bibinfo {author}
  {\bibfnamefont {Y.}~\bibnamefont {Zou}}, \bibinfo {author} {\bibfnamefont
  {J.~D.}\ \bibnamefont {Hidary}}, \bibinfo {author} {\bibfnamefont
  {G.}~\bibnamefont {Vidal}},\ and\ \bibinfo {author} {\bibfnamefont
  {S.}~\bibnamefont {Leichenauer}},\ }\bibfield  {title} {\bibinfo {title}
  {Tensornetwork: A library for physics and machine learning},\ }\href
  {https://api.semanticscholar.org/CorpusID:146120898} {\bibfield  {journal}
  {\bibinfo  {journal} {ArXiv}\ }\textbf {\bibinfo {volume} {abs/1905.01330}}
  (\bibinfo {year} {2019})}\BibitemShut {NoStop}%
\bibitem [{\citenamefont {M{\"u}ller}(2009)}]{Mller2009TheBA}%
  \BibitemOpen
  \bibfield  {author} {\bibinfo {author} {\bibfnamefont {S.}~\bibnamefont
  {M{\"u}ller}},\ }\bibfield  {title} {\bibinfo {title} {The binomial approach
  to option valuation: Getting binomial trees into shape}\ }(\bibinfo {year}
  {2009})\BibitemShut {NoStop}%
\bibitem [{\citenamefont {Korn}\ and\ \citenamefont
  {M{\"u}ller}(2009)}]{Korn2009p1}%
  \BibitemOpen
  \bibfield  {author} {\bibinfo {author} {\bibfnamefont {R.}~\bibnamefont
  {Korn}}\ and\ \bibinfo {author} {\bibfnamefont {S.}~\bibnamefont
  {M{\"u}ller}},\ }\bibfield  {title} {\bibinfo {title} {The decoupling
  approach to binomial pricing of multi-asset options},\ }\href@noop {}
  {\bibfield  {journal} {\bibinfo  {journal} {The Journal of Computational
  Finance}\ }\textbf {\bibinfo {volume} {12}},\ \bibinfo {pages} {1} (\bibinfo
  {year} {2009})}\BibitemShut {NoStop}%
\bibitem [{\citenamefont {Korn}\ and\ \citenamefont
  {Müller}(2009)}]{Korn2009p2}%
  \BibitemOpen
  \bibfield  {author} {\bibinfo {author} {\bibfnamefont {R.}~\bibnamefont
  {Korn}}\ and\ \bibinfo {author} {\bibfnamefont {S.}~\bibnamefont {Müller}},\
  }\bibfield  {title} {\bibinfo {title} {Getting multi-dimensional trees into a
  new shape},\ }\href {https://doi.org/https://doi.org/10.1002/wilj.12}
  {\bibfield  {journal} {\bibinfo  {journal} {Wilmott Journal}\ }\textbf
  {\bibinfo {volume} {1}},\ \bibinfo {pages} {145} (\bibinfo {year} {2009})},\
  \Eprint
  {https://arxiv.org/abs/https://onlinelibrary.wiley.com/doi/pdf/10.1002/wilj.12}
  {https://onlinelibrary.wiley.com/doi/pdf/10.1002/wilj.12} \BibitemShut
  {NoStop}%
\bibitem [{\citenamefont {Ito}(1944)}]{ito}%
  \BibitemOpen
  \bibfield  {author} {\bibinfo {author} {\bibfnamefont {K.}~\bibnamefont
  {Ito}},\ }\href@noop {} {\bibinfo {title} {Stochastic integral}} (\bibinfo
  {year} {1944})\BibitemShut {NoStop}%
\bibitem [{\citenamefont {Jacobs}(2010)}]{Jacobs_2010}%
  \BibitemOpen
  \bibfield  {author} {\bibinfo {author} {\bibfnamefont {K.}~\bibnamefont
  {Jacobs}},\ }\href@noop {} {\emph {\bibinfo {title} {Stochastic Processes for
  Physicists: Understanding Noisy Systems}}}\ (\bibinfo  {publisher} {Cambridge
  University Press},\ \bibinfo {year} {2010})\BibitemShut {NoStop}%
\end{thebibliography}
\end{document}